\documentclass[aps,floatfix,onecolumn,a4paper,noshowpacs, showkeys, nofootinbib,superscriptaddress,11pt]{revtex4}
\usepackage{float,xcolor,upgreek,yfonts,tikz}
\usepackage{color}
\usepackage{xcolor}
\usepackage{amsmath,amssymb,upgreek}

\usepackage{subcaption}
\usepackage{physics}
\usepackage[dvipsnames]{xcolor}

\usepackage[colorlinks,hyperindex,unicode]{hyperref}
\definecolor{green1}{RGB}{0,128,0} 
\hypersetup{%
  colorlinks = true,
  linkcolor  = blue,
  citecolor = cyan,
}
\usepackage{bookmark,textgreek}

\newcommand\orcidkelvin{{\href{https://orcid.org/0000-0002-3822-9818}{\orcidicon}}}

\newcommand\orcidrogerio{{\href{https://orcid.org/0000-0001-7848-5472}{\orcidicon}}}

\newcommand\orcidsantiago{{\href{https://orcid.org/0000-0002-2525-0386}{\orcidicon}}}

\newcommand{\orcidicon}{%
	\begin{tikzpicture}
	\draw[lime, fill=lime] (0,0)
		circle [radius=0.16]
		node[white] {{\fontfamily{qag}\selectfont \tiny ID}};
	\draw[white, fill=white] (-0.0625,0.095)
		circle [radius=0.007];
	\end{tikzpicture}	\hspace{-2mm}
}

\usepackage{graphicx}
\usepackage{dcolumn}
\usepackage{bm}
\usepackage{caption}
\usepackage{subcaption}
\usepackage{comment} 
\usepackage{float}
\usepackage{booktabs}
\begin{document}

\title{Light Deflection due to Spinoptic Effects in Parametrized and Spherically Symmetric Hairy Black Holes}

\author{Kelvin S. Alves\orcidkelvin}
\affiliation{Departament of Physics, São Paulo State University (UNESP), School of
Engineering and Sciences, Guaratinguetá, Brazil}
\email{kelvin.santos@unesp.br}

\author{Rogerio T. Cavalcanti\orcidrogerio}
\affiliation{Departament of Physics, São Paulo State University (UNESP), School of
Engineering and Sciences, Guaratinguetá, Brazil}
\affiliation{Institute of Mathematics and Statistics, Rio de Janeiro State University (UERJ),  Rio de Janeiro, Brazil}
\email{rogerio.cavalcanti@ime.uerj.br}

\author{Santiago E. Perez Bergliaffa\orcidsantiago}
\affiliation{Institute of Physics, Rio de Janeiro State University (UERJ), Rio de Janeiro, Brazil}
\email{santiago.bergliaffa@uerj.br}

\medbreak
\begin{abstract} 
In the standard geometric optics approximation, null rays propagating  
{in a spherically symmetric black hole background}
follow planar geodesics. This picture changes, however, when the helicity-dependent effects of light are incorporated into the dynamics. Specifically, the interaction between the helicity of light and the spacetime curvature induces a significant angular deflection out of the geodesic plane. In this paper, we employ the spinoptics formalism to study light deflection due to the helicity-curvature interaction in two spherically symmetric {geometries}: the Rezzolla--Zhidenko (RZ) parametrized metric, and a hairy regular black hole solution obtained via gravitational decoupling. Our results reveal clear imprints of both the RZ parametrization coefficients and the hairy black hole parameter on the deflection angle. Furthermore, we assess the viability of using the RZ parametrization to mimic the regular hairy black hole, discussing the validity and limitations of such an approximation.
\end{abstract}



\maketitle

\section{Introduction}
\label{sec0}

One of the most effective ways to obtain information about astrophysical objects is through the detection of electromagnetic radiation emitted by them or by their surroundings \cite{abbott2017multi}. This type of observation, fundamental to the development of modern astrophysics, has also gained relevance in the investigation of black hole physics. An example is the construction of images of the black holes $M87^{\star}$ and Sagittarius $A^{\star}$, as reported by the \textit{Event Horizon Telescope} collaboration \cite{akiyama2019first, EventHorizonTelescope:2022wkp}. The images were constructed from the electromagnetic radiation emitted by the accretion disks of the black holes, combining data obtained by a global network of radio telescopes. The electromagnetic waves travel long distances before reaching detectors. Hence, studying their propagation in curved spacetime proves to be essential not only from a theoretical standpoint but also for the interpretation and analysis of observational data. 

From the perspective of the geometric optics approximation, the propagation of electromagnetic waves in curved spacetimes is well- understood. This approach aims to find high-frequency asymptotic solutions to Maxwell's equations \cite{sachs1961gravitational,robinson1961null,kristian1966observations}. Ensuring the validity and covariance of the approximation requires that conditions involving the characteristic length scales of the problem be satisfied. The principal results may be summarized as follows \cite{Misner:1973prb}:
\begin{enumerate}
    \item The wave vector $\ell^{\mu}$ is null and tangent to a null geodesic;
    \item The polarization vector is orthogonal to the wave vector $\ell^{\mu}$ and is parallely-transported along the geodesic;
    \item Let $a$ be the scalar amplitude. The conservation equation $\nabla_{\mu}(a \ell^{\mu}) = 0$ ensures that the number of photons in a beam remains constant along the propagation.
\end{enumerate}
The geometric optics approximation, however, does not capture the gravitational spin-Hall effect, which stems from the helicity--curvature interaction: depending on helicity, {light rays}
follows slightly different paths~\cite{Oancea:2020khc}. Geometric optics is insensitive to helicity, so modifications are required. This leads to the \emph{spinoptics approximation}, valid for high but finite frequencies, with corrections of order $\mathcal{O}(1/\omega)$~\cite{Dahal:2023dsz}. The main result is that {light rays}
in curved spacetime remains null but generally does not follow geodesics~\cite{Andersson:2023bvw}. Such formalism has received extensive attention in the literature. Its application to electromagnetic wave propagation in curved, stationary, and arbitrary spacetimes is discussed in Refs.~\cite{Frolov:2011mh,Yoo:2012vv,Frolov:2020uhn,Dahal:2022gop}. Applications include light scattering by a rotating black hole~\cite{Frolov:2012zn}, light propagation in Kerr spacetime~\cite{Dahal:2023ncl}, and gravitationally induced birefringence~\cite{Murk:2024qgj}. More recently, the formalism has been extended to gravitational wave propagation~\cite{Cusin:2019rmt,Dahal:2021qel,Kubota:2023dlz}.

Whether through electromagnetic radiation or gravitational waves \cite{abbott2017gw170817}, modern astrophysics has enabled tests of fundamental aspects of General Relativity and its extensions, reaching previously inaccessible regimes \cite{Isi:2019asy, Ezquiaga:2020dao}. However, {due to the large number of modified theories of gravity}, testing each of them individually becomes impractical. {It is therefore useful to adopt general parametrization of the most relevant solutions, with parameters that can be determined observationally, in order to 
validate or rule out alternative theories of gravity.} %
%
%
%
%
Several parametrizations of 
{geometries describing the exterior of}
compact objects have been proposed in the literature. Prominent examples include the parametrized post-Newtonian (PPN) approach for compact binary systems~\cite{Will:2005va}. and the well-known Johannsen-Psaltis parametrization for rotating black holes deviating from the Kerr metric~\cite{Johannsen:2011dh}. The latter consists of a Taylor series in powers of $M/r$, where $r$ is the radial coordinate and $M$ is the black hole mass. An interesting parametrization approach for spherically symmetric black holes was proposed by Rezzolla and Zhidenko (RZ) in Ref.~\cite{Rezzolla:2014mua}. Using a continued fraction expansion with compact radial coordinates, it {describes}
deviations from General Relativity and achieves high accuracy with few expansion coefficients~\cite{DeLaurentis:2017dny, Konoplya:2020hyk}. By contrast, the Johannsen-Psaltis parametrization, for example, requires higher-order terms for convergence.

{An interesting alternative to the parametrized metric is offered by solutions obtained via gravitational decoupling (GD), a method introduced by Ovalle in Ref.~\cite{Ovalle:2017fgl}. GD extends known solutions by incorporating corrections to the energy–momentum tensor due to an extra matter source, fundamental fields, or geometric modifications.} This formalism has been widely applied in the literature to generate new black hole solutions \cite{Ovalle:2023ref,Ovalle:2020kpd,Ovalle:2023vvu,Hua:2025qwu}, physically realistic compact stellar configurations \cite{Estrada:2019aeh, Gabbanelli:2019txr, Leon:2023nbj,Ramos:2021drk, Sharif:2023ecm, Rincon:2019jal, Morales:2018urp}, as well as configurations exhibiting pressure anisotropies \cite{Gabbanelli:2018bhs, PerezGraterol:2018eut, Heras:2018cpz, Torres:2019mee, Contreras:2019iwm, Tello-Ortiz:2021kxg, Andrade:2023wux, Bamba:2023wok, Maurya:2023uiy, Iqbal:2025xqf, Tello-Ortiz:2023poi, Contreras:2021xkf}. 
Potentially observable properties of black hole solutions {obtained via the GD method}
have been discussed in Refs.~\cite{dePaiva:2025eux, Guimaraes:2025jsh, Cavalcanti:2022adb, Cavalcanti:2022cga, Ditta:2023arv, Wang:2025hla}. {In order to test the compatibility of the two approaches, we apply the Rezzolla–-Zhidenko (RZ) parametrization to a regular hairy black hole solution obtained via the GD method \cite{Ovalle:2023ref}.} To this end, we explicitly compute the expansion coefficients up to first order and assess their accuracy in describing the exact solution by calculating the shadow radius. We further investigate how the RZ coefficients behave as functions of the 
additional parameter of the hairy black hole.

After establishing the effect of interest, namely, spinoptics effects on light-ray propagation, and specifying the spacetime metric modelled by the RZ parametrization and the hairy regular GD black hole, we investigate how the additional parameters of these black holes influence light-ray trajectories when the interaction between spacetime curvature and photon helicity is taken into account. To this end, we first extend the formalism developed in Refs. \cite{Frolov:2024ebe, Frolov:2024olb} to generalized spherically symmetric spacetimes, in which $g_{rr}g_{tt}\neq 1$. {Our analysis reveals significant modifications that could potentially be tested in future observations, either through comparison with spinoptics results in a Schwarzschild background or within alternative theories of gravity that lack helicity-curvature interactions. We discuss these results and illustrate them with plots and simulations of light ray orbits incorporating helicity-curvature corrections.}

This work is organized as follows. Section~\ref{sec_2} introduces the GD formalism and reproduces the derivation of the regular hairy black hole solution. Section~\ref{sec3} presents the RZ parametrization and derives its expansion coefficients for the exact regular hairy solution, with particular attention to how the hair parameter affects the parametrization. Section~\ref{sec4} reviews the recently proposed spinoptics formalism. Section~\ref{sec5} applies the spinoptics formalism to both the RZ expansion and the regular hairy solution. Finally, Section~\ref{sec6} presents our conclusions. In Appendix \ref{app_null_tetrad}, we extend the parallel-propagated null tetrad to generalized spherically symmetric spacetimes.

\section{Regular Black Hole from Gravitational Decoupling}\label{sec_2}

This section briefly reproduces the derivation of a regular hairy black hole solution \cite{Ovalle:2023ref}, as obtained through the gravitational decoupling (GD) method \cite{Ovalle:2017fgl}. We begin with the Einstein field equations
\begin{align}
	G_{\mu\nu} = -k^{2} \tilde{T}_{\mu\nu},
\end{align}
where $G_{\mu\nu} = R_{\mu\nu} - \frac{1}{2}R g_{\mu\nu}$ is the Einstein's tensor and $\tilde{T}_{\mu\nu}$ denotes a generic energy-momentum tensor. The starting point is to assume that $\tilde{T}_{\mu\nu}$ can be decomposed as the sum of two distinct energy-momentum tensor contributions, namely
\begin{align}
	\tilde{T}_{\mu\nu} = T_{\mu\nu} + \Theta_{\mu\nu}.
\end{align}
In this context, $T_{\mu\nu}$ denotes an energy–momentum tensor associated with a solution of the Einstein field equations. The tensor $\Theta_{\mu\nu}$, on the other hand, is not determined a priori. It can originate from fundamental or effective fields, or from a gravitational sector beyond General Relativity. Given that the Einstein tensor satisfies the Bianchi identity, the total energy–momentum tensor must fulfill the conservation equation $\nabla_{\mu}\tilde{T}^{\mu\nu}= 0$.

Assuming a spherically symmetric spacetime, the metric in spherical polar coordinates is given by
\begin{align}\label{general_spherical_metric}
	{\rm d}s^{2} = -e^{\nu(r)} {\rm d}t^{2} + e^{\lambda(r)}{\rm d}r^{2} + r^{2} {\rm d}\theta^{2} + r^{2}\sin^{2}\theta {\rm d}\phi^{2}.
\end{align}
The substitution of the metric \eqref{general_spherical_metric} into Einstein's equations yields
\begin{align}\label{Eisntein_equations_general_}
	&k^{2}\left(T_{0}^{\phantom{i}0} + \Theta_{0}^{\phantom{i}0}\right) = -\frac{1}{r^{2}} - \left(\lambda' - \frac{1}{r}\right)\frac{e^{-\lambda}}{r}, \nonumber\\
	&k^{2}\left(T_{1}^{\phantom{i}1} + \Theta_{1}^{\phantom{i}1}\right) = -\frac{1}{r^{2}} + \left(\nu' + \frac{1}{r}\right)\frac{e^{-\lambda}}{r}, \nonumber \\
	&k^{2}\left(T_{2}^{\phantom{i}2} + \Theta_{2}^{\phantom{i}2}\right) = \left[2r\nu'' + r\left(\nu'\right)^{2} - \left(r\lambda' - 2\right)\nu' - 2\lambda'\right]\frac{e^{-\lambda}}{4r},
\end{align}
where the prime $(')$ denotes differentiation with respect to $r$, and the components satisfy $T_{2}^{\phantom{i}2} = T_{3}^{\phantom{i}3}$ due to spherical symmetry. Assuming a perfect fluid model for $T_{\mu\nu}$ and $\Theta_{\mu\nu}$, the effective density, radial pressure, and tangential pressure can be identified, respectively, as
\begin{align}
	&\tilde{\rho} \equiv T_{0}^{\phantom{i}0} + \Theta_{0}^{\phantom{i}0} \nonumber \\
	&\tilde{p}_r \equiv -T_{1}^{\phantom{i}1} - \Theta_{1}^{\phantom{i}1} \nonumber \\
	& \tilde{p}_{t} \equiv -T_{2}^{\phantom{i}2} - \Theta_{2}^{\phantom{i}2}.
 \end{align}
In general, $\Pi \equiv \tilde{p}_r - \tilde{p}t \neq 0$, implying that the GD method is capable of generating anisotropic fluid solutions, where the anisotropy originates from the presence of the source $\Theta_{\mu\nu}$.

Now, assuming a vanishing additional source, $\Theta_{\mu\nu} = 0$, and that the solution of Einstein's equations for the energy-momentum tensor $T_{\mu\nu}$, referred to as the seed solution, is given by
\begin{align}\label{seed_general_metric}
	{\rm d}s^{2}_{\rm seed} = -e^{\xi(r)} {\rm d}t^{2} + e^{\mu(r)}{\rm d}r^{2} + r^{2} {\rm d}\theta^{2} + r^{2}\sin^{2}\theta {\rm d}\phi^{2},
\end{align}
the Einstein equations become
	\begin{align}\label{Einstein_equations_seed_solution_}
		&k^{2}T_{0}^{\phantom{i}0} = -\frac{1}{r^{2}} - \left(\mu' - \frac{1}{r}\right)\frac{e^{-\mu}}{r}, \nonumber\\
		&k^{2}T_{1}^{\phantom{i}1} = -\frac{1}{r^{2}} + \left(\xi' + \frac{1}{r}\right)\frac{e^{-\mu}}{r}, \nonumber \\
		&k^{2}T_{2}^{\phantom{i}2}  = \left[2r\xi'' + r\left(\xi'\right)^{2} - \left(r\mu' - 2\right)\xi' - 2\mu'\right]\frac{e^{-\mu}}{4r}.
	\end{align}

To take the source $\Theta_{\mu\nu}$ back into account, it is assumed that it is encoded in the metric \eqref{seed_general_metric} via a geometric deformation of the radial and temporal components. This deformation, described by two new functions $f = f(r)$ and $g = g(r)$, can be expressed as
\begin{align}
&\xi \rightarrow \nu = \xi + \alpha g, 
&e^{-\mu} \rightarrow e^{-\lambda} = e^{-\mu} + \alpha f. \label{general_deformation_radial}
\end{align}
The constant parameter $\alpha$ indicates the deviation from the seed solution. The essence of the decoupling method lies in the fact that substituting Eq. \eqref{general_deformation_radial} into the Einstein equations \eqref{Eisntein_equations_general_} yields two independent systems. The first corresponds to the Einstein equations for the source $T_{\mu\nu}$, given by Eqs. \eqref{Einstein_equations_seed_solution_}. The second comprises the field equations for the additional source $\Theta_{\mu\nu}$, which take the form
	\begin{align}
		k^{2}\Theta_{0}^{\phantom{i}0} &= -\frac{\alpha}{r}\left(f'+\frac{f}{r}\right),\label{quasi_einstein_equation_one} \\
		k^{2}\Theta_{1}^{\phantom{i}1} + \alpha \mathcal{J}_{1} &= -\alpha f\left(\frac{\nu'r + 1}{r^{2}}\right), \label{quasi_einstein_equation_two}\\
		k^{2}\Theta_{2}^{\phantom{i}2} + \alpha \mathcal{J}_{2} &= -\frac{\alpha f}{4r} \left(2r\nu'' + r\nu'^{2} + 2\nu'\right) - \frac{\alpha f'}{4r}\left(r\nu' + 2\right), \label{quasi_einstein_equation_three}
	\end{align}
with
\begin{align}
	\mathcal{J}_{1} &= \frac{e^{-\mu}g'}{r} \\
	\mathcal{J}_{2} &= \frac{e^{-\mu}}{4}\left(2g'' + \alpha g'^{2} + \frac{2g'}{r} + 2\xi'g' - \mu g'\right).
\end{align}
This system of equations is also known as the quasi-Einstein equations. Note that, when $\displaystyle\lim_{\alpha \to 0} \Theta_{\mu\nu} =0$ the seed solution is recovered. 

Assuming the Schwarzschild solution as the seed metric, i.e., $e^{\xi(r)} = e^{-\mu(r)} = 1 - \frac{2M}{r}$, and imposing $e^{\nu(r_h)} = e^{\lambda(r_h)} = 0$ to ensure that the causal horizon of the extended solution ($e^{-\lambda(r_h)} = 0$) coincides with its Killing horizon ($e^{\nu(r_h)} = 0$), where $r = r_h$ denotes the event horizon radius, the line element \eqref{general_spherical_metric} becomes
\begin{align}
	{\rm d}s^{2} = -\left(1-\frac{2M}{r}\right)h(r) {\rm d}t^{2} + \left(1-\frac{2M}{r}\right)^{-1} \frac{{\rm d}r^{2}}{h(r)} + r^{2} {\rm d}\theta^{2} + r^{2}\sin^{2}\theta {\rm d}\phi^{2},
\end{align}
with $h(r) = e^{\alpha g(r)}$, and $g(r)$ is a function to be determined.

Due to the additional degrees of freedom arising from the deformation functions, the field equations do not completely determine the solution, and additional constraints must be imposed. The choice made in \cite{Ovalle:2023ref} is to require that the additional source $\Theta_{\mu\nu}$ satisfies the weak energy condition, given by
\begin{align}
&\tilde{\rho} \geq 0,\label{WEC_one} \\
&\tilde{\rho} + \tilde{p}_{r} \geq 0, \\
&\tilde{\rho} + \tilde{p}_{\Theta} \geq 0,
\end{align}
with the energy density modeled as \cite{Ovalle:2023ref}
\begin{align}\label{energy_density_WEC}
k \tilde{\rho} = \frac{\alpha}{\ell^{2}}e^{-r/\ell}.
\end{align}
The parameter ${\ell}$ is a constant with dimensions of length, and $\alpha$ was introduced to recover the vacuum seed solution in the limit $\alpha \rightarrow 0$. After some manipulations, the regular hairy black hole metric can be obtained
\begin{align}
e^{\nu(r)} = e^{-\lambda(r)} = 1 + \frac{2M}{r} + \frac{e^{-\alpha r/M}}{r M}\left(2M^{2} + 2M\alpha r + \alpha^{2}r^{2}\right).
\end{align}
Here, $M = \alpha {\ell}$ is the ADM mass. Throughout this work, for convenience, we set $\alpha = \beta^{-1}$ and express radial coordinates in units of the Schwarzschild mass, so that $r/M \mapsto r$. Under this rescaling, the metric takes the form
\begin{align}\label{regular_metric_black_hole}
	{\rm d}s^{2} = -f(r,\beta) {\rm d}t^{2} + f(r,\beta)^{-1} {\rm d}r^{2}  + r^{2}{\rm d}\theta^{2}+r^{2}\sin^{2}\theta {\rm d}\phi^{2},
\end{align}
with
\begin{align}\label{functions_regular_metric}
	f(r,\beta) \equiv e^{\nu(r)} = e^{-\lambda(r)} = 1 -  \frac{2}{r}  +  \frac{2e^{-\frac{r}{\beta}}}{r}  \left( 1 + \frac{r}{\beta}  +\frac{r^2}{2\beta^2} \right). 
\end{align}
Notice that in the limits $\beta \to 0$ and $\beta \to \infty$, we recover the Schwarzschild and Minkowski spacetimes, respectively, since $\lim_{\beta\to 0} f(r,\beta)= 1-\frac{2}{r}$ and $\lim_{\beta\to \infty} f(r,\beta) = 1$.

The event horizon radius is determined by the condition $f(r_h, \beta) = 0$. Depending on the value of the parameter $\beta$, three distinct possibilities arise, separated by a critical value $\beta_{\text{crit}} \approx 0.3883945$. For $\beta < \beta_{\text{crit}}$, there are two horizons: a Cauchy horizon and the event horizon. For $\beta > \beta_{\text{crit}}$, no horizon exists, and the solution does not describe a black hole. The extremal configuration, corresponding to $\beta = \beta_{\text{crit}}$, features a single horizon.

To determine whether the solution is non-singular and to analyze the behavior of the spacetime curvature arising from the metric \eqref{regular_metric_black_hole}, we evaluate the Kretschmann scalar \cite{Henry:1999rm}, which is given by
\begin{align}
	K(r,\beta) =& \frac{48}{r^{6}} -8 \, {\left(\frac{12}{r^{6}} + \frac{1}{r^{2} \beta^{4}} + \frac{2}{r^{3} \beta^{3}} + \frac{6}{r^{4} \beta^{2}} + \frac{12}{r^{5} \beta}\right)} e^{\left(-\frac{r}{\beta}\right)} \nonumber \\&+ {\left(\frac{48}{r^{6}} + \frac{r^{2}}{\beta^{8}} + \frac{8}{\beta^{6}}+ \frac{16}{r \beta^{5}} + \frac{36}{r^{2} \beta^{4}} + \frac{64}{r^{3} \beta^{3}} + \frac{96}{r^{4} \beta^{2}} + \frac{96}{r^{5} \beta}\right)} e^{\left(-\frac{2 \, r}{\beta}\right)}. 
\end{align}
Note that the Kretschmann scalar $K$ corresponds to that of the Schwarzschild solution, corrected by terms arising from the black hole hair parameter. These additional terms are responsible for its regularity. This property can be observed by taking the limit $r \to 0$, approaching the black hole center.
\begin{align}
	\lim_{r \to 0} K = \frac{8}{3\beta^{6}}.
\end{align}
In this limit, the curvature will depend only on the parameter $\beta$, so that for $\beta \to 0$, the singularity of the Schwarzschild spacetime is recovered at the center of the black hole.
\begin{figure}[h!]
	\centering
	\includegraphics[width=0.6\linewidth]{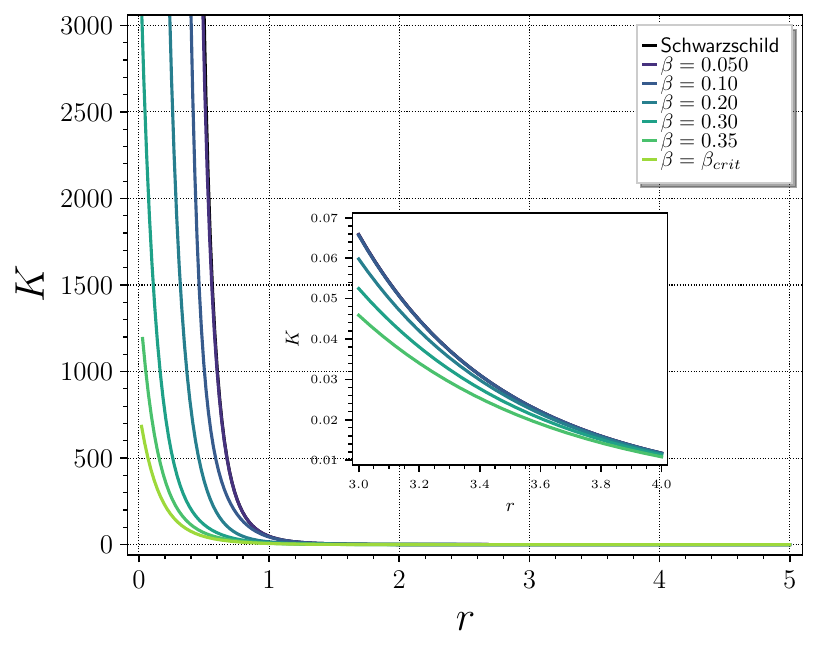}
	\caption{\footnotesize Kretschmann scalar of the hairy regular solution as a function of the radial distance for various values of the parameter $\beta$.}
	\label{fig:Kretschmann_Scalar}
\end{figure}
In Fig. \ref{fig:Kretschmann_Scalar}, the Kretschmann scalar is plotted as a function of the radial distance from the black hole for various values of the parameter $\beta$. It can be seen that the spacetime curvature is affected by the presence of the hair parameter. In particular, we observe that the curvature is smaller for larger values of $\beta$.

\section{Rezzolla--Zhidenko Parametrization for Hairy Black Holes}
\label{sec3}

This section focuses on the parametrized spherically symmetric black holes introduced by Rezzolla and Zhidenko (RZ) \cite{Rezzolla:2014mua}. In particular, we investigate whether it can fully describe the hairy solution obtained via gravitational decoupling in the previous section (see Eq.\eqref{regular_metric_black_hole}). To this end, we establish a relation between the coefficients of the parametrization and the parameter of the hairy solution. We begin with the RZ metric, given by
\begin{align}\label{parametrized_BH_solution}
	{\rm d}s^{2} = - N^{2}(r) {\rm d}t^{2} + \frac{B^{2}(r)}{N^{2}(r)}{\rm d}r^{2} + r^{2} {\rm d}\theta^{2} + r^{2}\sin^{2}\theta {\rm d}\phi^{2},
\end{align}
where $N^{2}(r)$ can be expressed as
\begin{align}\label{N_parametrized_metric_x}
    N^{2}(x) = xA(x),
\end{align}
with $ A(x)>0$ and $x \equiv 1 - \frac{r_{h}}{r}$. The dimensionless compact radial coordinate $x$ is defined on the interval $0 \leq x \leq 1$, where $x = 0$ ($r = r_h$) locates the event horizon and $x = 1$ ($r \to \infty$) corresponds to spatial infinity. The functions $A(x)$ and $B(x)$ are expressed in terms of the expansion parameters $\epsilon$, $a_i$, and $b_i$, with $i =0,1,2,\ldots,n$ and $n$ is the order of the expansion. These functions take the form
\begin{align}\label{def_A(x)}
A(x) = 1 - \epsilon (1-x) + (a_{0} - \epsilon) (1-x)^{2} + \tilde{A}(x) (1-x)^{3},
\end{align}
and
\begin{align}
B(x) = 1 + b_{0} (1 - x) + \tilde{B}(x)(1-x)^{2}.
\end{align}
Here, $\tilde{A}(x)$ and $\tilde{B}(x)$ are provided by continued fraction expansions derived via Padé approximation as follows
\begin{align}
	\tilde{A}(x) = \frac{a_{1}}{1 + \frac{a_{2}x}{1 + \frac{a_{3}x}{1 + \dots}}}, \qquad 	\tilde{B}(x) = \frac{b_{1}}{1 + \frac{b_{2}x}{1 + \frac{b_{3}x}{1 + \dots}}}.
\end{align}
These continued fraction expansions are determined by the behavior of the metric in the neighborhood of the horizon ($x \approx 0$) and at spatial infinity ($x \approx 1$). The dimensionless coefficients $a_i$ and $b_i$ are then fixed either by near-horizon phenomena or by matching to a specific metric. The parameter $\epsilon$ quantifies the shift in the horizon position of the parametrized solution relative to that of the Schwarzschild black hole, and takes the form \cite{Rezzolla:2014mua}
\begin{align}\label{epsilon_parametrization}
\epsilon = \frac{2 - r_h}{r_h}.
\end{align}

\begin{figure}[h!]
    \centering
       \includegraphics[width=0.48\linewidth]{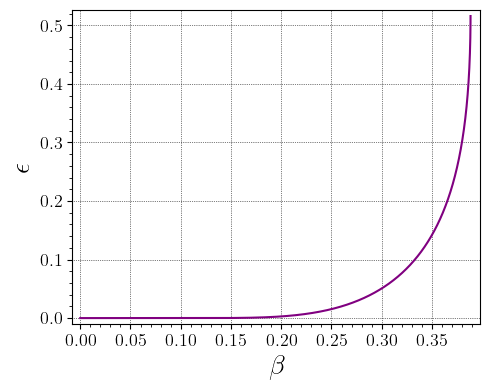}
    \caption{\footnotesize Relative deviation of the regular black hole horizon from the Schwarzschild horizon as a function of the deviation parameter $\epsilon$ and the hairy parameter $\beta$.}
    \label{fig:eps_regular}
\end{figure}

For the hairy solution described by Eqs. \eqref{regular_metric_black_hole} and \eqref{functions_regular_metric}, $N^{2}(r,\beta) = f(r,\beta)$ and $B(r)=1$. Figure \ref{fig:eps_regular} shows the deviation parameter $\epsilon$ as a function of the hairy parameter $\beta$. It is seen that the larger $\beta$, the more the horizon radius of the regular black hole deviates from that of the Schwarzschild black hole. This deviation reaches its maximum at the critical value $\beta_{\text{crit}} \approx 0.3883945$.
Expanding the metric function \eqref{functions_regular_metric} at spatial infinity and comparing it with the RZ parametrization shows that the zeroth-order expansion coefficient, $a_0$, vanishes for the regular hairy black hole \cite{DeLaurentis:2017dny}. Moreover, the first-order coefficient $a_1$ can be determined by comparing the behavior of the metric function \eqref{functions_regular_metric} near the horizon with that of the RZ metric in the same region. Following Ref. \cite{Konoplya:2022tvv}, we find
 \begin{align}\label{deriv_N2_}
    	\left.\frac{{\rm d} N^{2}(x)}{{\rm d}x}\right|_{x=0} = \left[A(x) + x\frac{{\rm d}A(x)}{{\rm d}x}\right]_{x=0} = A(0),
\end{align}
where
\begin{align}\label{deriv_N2_fbeta}
    \left.\frac{{\rm d} N^{2}(x)}{{\rm d}x}\right|_{x=0} = \left.\frac{{\rm d}f(r,\beta)}{{\rm d}r}\right|_{r=r_{h}} \left.\frac{{\rm d}r}{{\rm d}x}\right|_{x=0}=  \frac{2}{{r_h}} -{\left(\frac{2}{{r_h}} + \frac{{r_h}^{2}}{\beta^{3}} + \frac{{r_h}}{\beta^{2}} + \frac{2}{\beta}\right)} e^{\left(-\frac{{r_h}}{\beta}\right)}.
\end{align} 
From the equation \eqref{def_A(x)}, on the other hand, we have
\begin{align}\label{eq_A(0)}
    A(0) =  1 - 2\epsilon + a_0 + a_1 = 1 - 2\epsilon + a_1.
\end{align}
Consequently, by employing equations \eqref{epsilon_parametrization}, \eqref{deriv_N2_}, \eqref{deriv_N2_fbeta}, and \eqref{eq_A(0)}, we can express the coefficient $a_1$ for the regular hairy black hole solution in terms of the hair parameter and the horizon radius
\begin{align}\label{a1_parametrization}
    a_{1} = \frac{6}{{r_h}}- 3 -{\left(\frac{2}{{r_h}} + \frac{{r_h}^{2}}{\beta^{3}} + \frac{{r_h}}{\beta^{2}} + \frac{2}{\beta}\right)} e^{\left(-\frac{{r_h}}{\beta}\right)}.
\end{align}
\begin{figure}[h!]
    \includegraphics[width=0.54\linewidth]{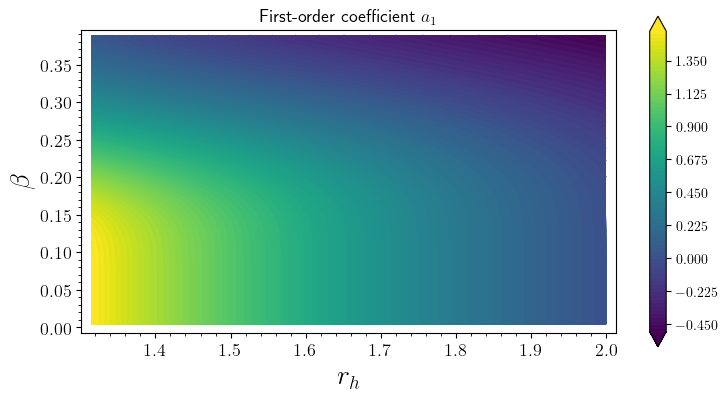}
    \includegraphics[width=0.44\linewidth]{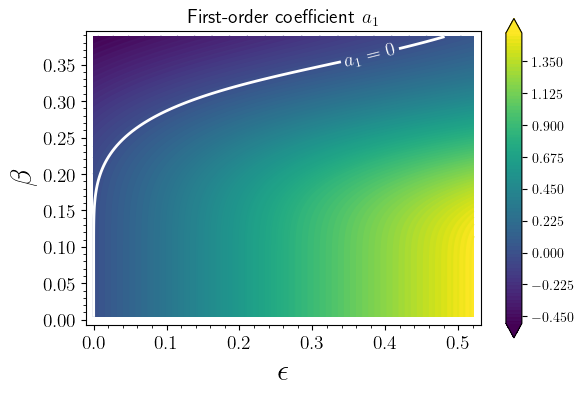}
    \caption{\footnotesize Behavior of the first-order coefficient $a_1$ of the continued fraction expansion in the Rezzolla–Zhidenko parametrization for the regular black hole solution. Left panel: $a_1$ as a function of the horizon radius $r_h$ and the hairy parameter $\beta$. Right panel: $a_1$ as a function of the deviation parameter $\epsilon$ and the hairy parameter $\beta$.}
    \label{fig:placeholder}
\end{figure}
The first-order coefficient $a_1$ of the continued fraction expansion in the RZ parametrization is shown in Fig. \ref{fig:placeholder}. In the left panel, we present $a_1$ for the regular black hole solution with hair as a function of the horizon radius $r_h$ and the hair parameter $\beta$. The coefficient decreases as either $\beta$ or $r_h$ increases. This behavior is opposite to that of the deviation parameter $\epsilon$, which increases as $\beta$ approaches its critical value. This contrast is more clearly seen in the right panel of Fig. \ref{fig:placeholder}.

To evaluate the accuracy of the first-order approximation in the RZ parametrization, we compute the black hole shadow radius, which constitutes a directly observable quantity. For this purpose, we use the parametrization expressed in terms of the coefficients \eqref{epsilon_parametrization} and \eqref{a1_parametrization} and compare the resulting value with the exact one derived from the solution \eqref{regular_metric_black_hole} and \eqref{functions_regular_metric} \cite{dePaiva:2025eux}. Following Ref. \cite{Konoplya:2020hyk}, we define the function
\begin{align}
P(x) = x(1-x)^{2}A\left(x\right),
\end{align}
where $A\left(x\right)$ was introduced in Eq. \eqref{def_A(x)}. For the parametrized solution, the shadow radius $R_{sh}$ is obtained from
\begin{align}
\frac{r_{h}^{2}}{R_{sh}^{2}} = P(x_{m}),
\end{align}
where $x_{m}$ is the coordinate at which $P(x)$ attains its maximum value, corresponding to the location of the photon sphere.
\begin{table}[h!]
\centering
\renewcommand{\arraystretch}{1.2}
\begin{tabular}{ccccc}
\hline\hline
$\beta$ & $\epsilon$ & $a_1$ & $R_{sh}$ & $E$ \\
\hline
$0$                & $0$              & $0$               & $5.1962$ & $0\%$       \\
$0.1$              & $\sim 10^{-7}$   & $\sim -10^{-6}$   & $5.1962$ & $0.0001\%$  \\
$0.2$              & $0.0028$         & $-0.0175$         & $5.2031$ & $0.1379\%$  \\
$0.3$              & $0.0507$         & $-0.1342$         & $5.1917$ & $0.2083\%$  \\
$\beta_{\mathrm{crit}}$ & $0.5212$    & $0.0429$          & $4.7347$ & $6.7940\%$  \\
\hline\hline
\end{tabular}
\caption{\footnotesize Rezzolla–Zhidenko parameters $\epsilon$ and $a_{1}$, shadow radius of the parametrized solution, and relative error between the shadow radius obtained from the exact hairy regular solution and its approximation via the Rezzolla–Zhidenko parametrized metric.}
\label{tab:parameters_parametrized_black_hole}
\end{table}
Table \ref{tab:parameters_parametrized_black_hole} lists the values of the parameters $\epsilon$ and $a_1$ for different values of $\beta$, along with the black hole shadow radius obtained from the parametrized solution and the relative error between the exact shadow radius and its RZ approximation. Note that increasing $\beta$ leads to larger values of $\epsilon$ and $a_1$, consistent with the trends observed in the plots. Furthermore, the accuracy of the first-order parametrization decreases with increasing $\beta$, due to the growing departure from the Schwarzschild solution. Consequently, higher-order terms in the expansion would be needed to achieve greater precision.

\section{Spinoptics Formalism}
\label{sec4}

The spinoptics approach is employed to study the propagation of high-frequency electromagnetic waves by incorporating into the ray dynamics the effects arising from the interaction between helicity and spacetime curvature. These modified ray equations are known as the spinoptics equations. This helicity–curvature coupling becomes particularly significant in the presence of strong gravitational fields, as will be introduced here and further explored in Section \ref{sec5}.

In this section, we derive the spinoptics equations for a high-frequency electromagnetic wave using the effective action approach proposed by Frolov \cite{Frolov:2024ebe}. We begin with the Maxwell action in curved spacetime,
\begin{align}\label{action}
\mathcal{I} = \frac{1}{8}\int F_{\mu\nu}\bar{F}^{\mu\nu}\sqrt{-g} \: {\rm d}^4x,
\end{align}
where $F_{\mu\nu} = \nabla_{\mu}A_{\nu} - \nabla_{\nu}A_{\mu}$ is the complex Maxwell tensor, $\bar{F}^{\mu\nu}$ denotes its complex conjugate\footnote{Throughout this work, the complex conjugate of a quantity $A$ is denoted by $\bar{A}$.}, and $A_{\mu}$ is a complex vector potential. Adopting a WKB ansatz for the vector potential,
\begin{align}\label{ansatz_A}
A_{\mu} = a M_{\mu} e^{i\omega S},
\end{align}
where $a$ and $S$ are real scalar functions representing the amplitude and phase, respectively, and $M_{\mu}$ is a complex null vector satisfying the normalization conditions
\begin{align}\label{normalization_rule}
M^{\mu}M_{\mu} = \bar{M}^{\mu}\bar{M}_{\mu} = 0, \qquad M^{\mu}\bar{M}_{\mu} = 1, \qquad \bar{M}^{\mu}\partial_{\mu}S = 0.
\end{align}
Applying the ansatz in Eq. \eqref{ansatz_A} to the electromagnetic action \eqref{action} and imposing the normalization conditions in eq. \eqref{normalization_rule} yields
\begin{align}\label{eletromagnetic_WKB_action}
		\mathcal{I} = \frac{\omega^{2}}{2}\int\left\{a^{2}\left [\frac{1}{2}  (\nabla S)^{2}
		- \frac{1}{\omega}B^{\mu}\partial_{\mu}S 
		\right]\right\} \sqrt{-g}\: \mathrm{d}^{4}x ,
\end{align}
where $B_{\mu} = i\bar{M}^{\nu}\nabla_{\mu}M_{\nu}$ is a real vector. The spinoptics effective action can be identified from Eq. \eqref{eletromagnetic_WKB_action} as
\begin{align}\label{effetive_action_spinoptics}
	I = \int\left\{a^{2}\left [\frac{1}{2}  (\nabla S)^{2}
	- \frac{1}{\omega}B^{\mu}\partial_{\mu}S 
	\right]+ \Lambda\right\} \sqrt{-g} \:\mathrm{d}^{4}x.
\end{align}
where $\mathcal{I} = \omega^{2} I/2$. The two actions thus differ by a constant factor, which does not affect the equations of motion. The term $\Lambda$ encapsulates the constraints incorporated in deriving the spinoptics approximation and is given by 
\begin{align}
	\Lambda = \frac{1}{2}\bar{\lambda}_{1} M_{\mu}M^{\mu} + \frac{1}{2}\lambda_{1}\bar{M}_{\mu}\bar{M}^{\mu} + \lambda_{2}(M_{\mu}\bar{M}^{\mu}-1) + \bar{\lambda}_{3}M^{\mu}\partial_{\mu}S + \lambda_{3}\bar{M}^{\mu}\partial_{\mu}S,
\end{align}
where $\lambda_{1}$ and $\lambda_{3}$ are complex constants, and $\lambda_{2}$ is a real constant, and act as Lagrange multipliers.

From the variation 
of the effective action 
with respect to $a$, a Hamilton–Jacobi-type condition is obtained:
\begin{align}\label{Hamiltonian_Hamilton_Jacob}
H =\frac{1}{2} (\nabla S)^{2} - \frac{1}{\omega}B^{\mu}\partial_{\mu}S = 0,
\end{align}
while variation with respect to $S$ gives the conservation law
\begin{align}\label{eq_conservacao}
\nabla_{\mu}J^{\mu}=0,
\end{align}
where the current $J^\mu$ is defined as
\begin{align}
J^{\mu} = a^{2}\left(\partial^{\mu}S - \frac{1}{\omega}B^{\mu}\right) + \bar{\lambda}{3}M^{\mu} + \lambda{3}\bar{M}^{\mu}.
\end{align}
Finally, variation with respect to $M_{\mu}$, together with the invariance of $M^{\mu}$ under the transformation $M^{\mu} \rightarrow e^{i\psi}M^{\mu}$, allows us to impose the condition $B^{\mu}\partial_{\mu}S=0$. This leads to the dynamical equation
\begin{align}\label{dinamical_equation_}
	i\frac{a^{2}}{\omega}\partial^{\nu}S\nabla_{\nu}M_{\mu} = \lambda_{3}\partial_{\mu}S.
\end{align}
Equation \eqref{Hamiltonian_Hamilton_Jacob} can be identified as the Hamilton–Jacobi equation for a particle with four-momentum $p_{\mu} = \partial_{\mu}S$ and determines the trajectories of high-frequency electromagnetic waves
\begin{align}\label{spinoptic_hamilton_jacobi}
	H(p,x) = \frac{1}{2} g^{\mu\nu} p_{\mu}p_{\nu} - \frac{1}{\omega} B^{\mu}p_{\mu}.
\end{align}
Note that in the geometric optics limit, $\omega \to \infty$, Eq. \eqref{spinoptic_hamilton_jacobi} reduces to 
the Hamiltonian of geometric optics. 

The Hamilton equations that follow from Eq. \eqref{spinoptic_hamilton_jacobi} are
\begin{align}\label{hamilton_eqs_}
	\frac{\partial H}{\partial p_{\mu}} &= p_{\mu} - \frac{1}{\omega} B_{\mu} = \frac{{\rm d}x^{\mu}}{{\rm d}\lambda},\nonumber \\
	-\frac{\partial H}{\partial x^{\mu}} &= \frac{{\rm d}p_{\mu}}{{\rm d}x^{\mu}}.
\end{align}

{The above equations can be expressed more conveniently in a null tetrad basis. Indeed, let $\{\vb*{\ell}, \vb*{n}, \vb*{m}, \vb*{\bar{m}}\}$ be a null basis of the spacetime $(\mathcal{M},{g})$, satisfying the normalization conditions $\ell^\mu n_\mu=-1$ and $m^\mu\bar{m}_\mu=1$ with all other scalar products vanishing, and where $\vb*{\bar{m}}$ denotes the complex conjugate of $\vb*{m}$. Using these normalization conditions, one can show that}
\[
\epsilon^{\mu\nu\alpha\beta} \ell_{\mu} n_{\nu} m_{\alpha} \bar{m}_{\beta} = i\sigma,
\]
where $\epsilon^{\mu\nu\alpha\beta}$ is the Levi-Civita tensor and
\begin{align}
\sigma = \begin{cases}
 1, & \text{right-handed tetrad},\\
-1, & \text{left-handed tetrad}.
\end{cases}
\end{align}
The null ray $\vb*{\ell}$ can be interpreted as the four-vector describing the trajectory of a light ray interacting with spacetime curvature. Using Eq. \eqref{hamilton_eqs_}, it can be expressed as $\ell^{\mu} = p^{\mu} - \frac{1}{\omega}B^{\mu}$.  The second equation in \eqref{hamilton_eqs_} yields
\begin{align}\label{pre_geodesic_equation}
	D\ell^{\mu} = \ell^{\nu}\nabla_{\nu}\ell^{\mu} = w^{\mu} = \frac{1}{\omega} K^{\mu}_{\phantom{i}\nu}\ell^{\nu},
\end{align}
where $K^{\mu}_{\phantom{i}\nu} = \nabla_{\mu}B_{\nu} - \nabla_{\nu}B_{\mu}$ is an antisymmetric tensor. Moreover, $\ell^{\mu}\nabla_{\mu}(\vb*{\ell}^{2})=0$, which implies that if $\vb*{\ell}$ vanishes at one point along the trajectory, it vanishes everywhere. It is worth noting that $\vb*{\ell}$ is not a geodesic vector, but becomes one in the limit $\omega \to \infty$.

The choice of the null complex tetrad is constrained by the requirement that the same limit be recovered when its vectors are propagated along the null ray $\vb*{\ell}$. Accordingly, we have
\begin{align}
	D\ell^{\mu} &= \frac{1}{\omega}(\bar{\kappa}m^{\mu}+\kappa\bar{m}^{\mu}), \label{propagation_l}\\
	Dn^{\mu} &= 0, \label{propagation_n}\\
	Dm^{\mu} &= \frac{1}{\omega}\kappa n^{\mu}, \label{propagation_m}
\end{align}
where $\kappa$ are the complex coefficients of $w^{\mu}$ when expressed in the tetrad basis. Using the set of vectors defined above, the spinoptics equations, corresponding to eqs. \eqref{eq_conservacao}, \eqref{dinamical_equation_}, \eqref{propagation_l}, \eqref{propagation_n}, and \eqref{propagation_m}, can be written as
\begin{align}
	D\ell^{\mu} &= w^{\mu}, \label{eq_l_geodesic}\\
	Dn^{\mu} &= 0,\\
	Dm^{\mu} &= \frac{\sigma}{\omega}\kappa n^{\mu},\\
	\nabla_{\mu}(a^{2}\ell^{\mu}) &= 0. \label{conversation_ray}
\end{align}
Here 
\begin{align}\label{w_kappa}
w^{\mu} = \frac{\sigma}{\omega}\left(\bar{\kappa}m^{\mu} + \kappa\bar{m}^{\mu}\right) \quad \text{and} \quad \kappa = iR_{\mu\nu\alpha\beta}m^{\mu}\ell^{\nu}m^{\alpha}\bar{m}^{\beta}.
\end{align}
Equation \eqref{conversation_ray} indicates that the number of photons in a beam is conserved.

\section{Spinoptics for Parametrized and Hairy Black Holes}
\label{sec5}

The spinoptics formalism has previously been applied to the Schwarzschild black hole \cite{Frolov:2024olb} and the Kerr black hole \cite{Frolov:2025bva}. In this section, we apply it to study light propagation in the vicinity of the parametrized black holes introduced in Section \ref{sec3} and the hairy regular solution from Section \ref{sec_2}. To do so, we first construct a null tetrad basis that is parallel-propagated along the geodesic in the high-frequency limit $\omega \to \infty$, following the approach of Ref. \cite{Frolov:2024ebe}, which uses spacetime symmetries to obtain the null basis and then transforms it into a parallel-propagated one. In fact, we slightly generalize the null tetrad basis used in Frolov’s paper, which is valid for spherically symmetric spacetimes with a metric of the form $g = -f(r){\rm d}t^2 + \dfrac{1}{f(r)}{\rm d}r^2 + r^2{\rm d}\theta^2 + r^2\sin^2\theta {\rm d}\phi^2$. However, for the Rezzolla–Zhidenko metric, as shown in eq. \eqref{parametrized_BH_solution}, when $B^2(r)\neq 1$ the metric does not take the form assumed in \cite{Frolov:2024ebe}. The difference in the null tetrad construction arises because the equation $\nabla_\lambda h_{\mu\nu} = \xi_\mu g_{\nu\lambda} - \xi_\nu g_{\mu\lambda}$, whose solution is the conformal Killing–Yano two-form, imposes the condition $g_{tt}g_{rr}=c$ (with $c$ an integration constant). This condition does not hold for the RZ metric \eqref{parametrized_BH_solution}; consequently, no conformal Killing–Yano two-form exists. Nevertheless, the conformal Killing–Yano two-form and its dual (the Killing–Yano two-form) are used only to construct tensors that project vectors onto orthogonal directions. We thus use, instead, a two-form for projection that is not a conformal Killing–Yano tensor. The explicit construction of generalized null tetrad parallel-propagated along the geodesic is given in Appendix \ref{app_null_tetrad}.

\subsection{Parametrized Black Holes}

For practical reasons, we must first truncate the parametrized metric of Eq. \eqref{parametrized_BH_solution}. Following Refs. \cite{Konoplya:2022tvv, Toshmatov:2023anz, Toshmatov:2022kim}, we assume that the geometry is determined by only three parametrization coefficients: $\epsilon$, $a_1$, and $b_1$. This simplification reduces the complexity of the calculations while retaining good, {as discussed in Ref. \cite{Konoplya:2020hyk}}. Accordingly, we write the metric functions in \eqref{parametrized_BH_solution} as
\begin{align}\label{N2_first_order}
    N^{2}\left(r\right) = \left(1 - \frac{r_h}{r}\right)\left(1 - \epsilon\frac{r_h}{r} - \epsilon\frac{r_{h}^{2}}{r^{2}} + a_1\frac{r_{h}^{3}}{r^{3}} \right), 
\end{align}
\begin{align}\label{B2_first_order}
    B^{2}(r) = \left(1 + \frac{r_{h}^{2} b_{1}}{r^{2}}\right)^{2}.
\end{align}
Throughout this section, for convenience, we express radial coordinates in units of the horizon radius by rescaling $2r/r_h \mapsto r$. Under this rescaling, the metric functions take the form:
\begin{align}\label{N2_first_order_1}
    N^{2}\left(r\right) \equiv N^2 = \left(1 - \frac{2}{r}\right)\left(1 - \frac{2\epsilon}{r} - \frac{4\epsilon}{r^{2}} + \frac{8a_1}{r^{3}} \right), 
\end{align}
\begin{align}\label{B2_first_order_1}
    B^{2}(r) \equiv B^2= \left(1 + \frac{4b_{1}}{r^{2}}\right)^{2}.
\end{align}
When $\epsilon = a_1 = b_1 = 0$, the Schwarzschild solution is recovered. The coefficients $a_0$ and $b_0$ vanish to ensure consistency with General Relativity at first order in the post-Newtonian expansion \cite{Toshmatov:2021hdy}.

We now apply the formalism from the previous section, together with the null tetrad adapted to the parametrized spherically symmetric spacetime introduced in Appendix \ref{app_null_tetrad}. To this end, consider a null geodesic $x^{\mu}(\lambda)$ parametrized by an affine parameter $\lambda$. %
Using equation \eqref{w_kappa} and the null tetrad given by eqs. \eqref{tetrad_l}–\eqref{tetrad_n}, we obtain an expression for $\kappa$ for the parametrized black hole.
\begin{align}\label{kappa_parametrized}
    \kappa = \frac{\mathcal{A}}{2 L r^{4}\, \left(r^{2} + 4b_{1} \right)^{3}},
\end{align}
where
\begin{align}
   \mathcal{A} = &\sqrt{2}i L^{3} {\left[3   r^{5} - 20  r^{3} - 4 \, b_{1} \, r {\left(r^{2} + 4  \right)} \right]}  \Phi \epsilon \nonumber \\
    &+ \sqrt{2} i \, L{\left[-8  E^{2}  {b_1} r^{6} + 16  L^{2} {b_1} r^{4} + 3  L^{2} r^{5} + 64  L^{2} {b_1}^{3} - 16  L^{2} {a_1} {b_1} r \right.} \nonumber\\ 
    &{\left.+ 64  L^{2} {a_1} {b_1} 
     - 4 L^{2}\, {\left(5   {a_1} +   {b_1}\right)} r^{3} + 48 \, L^{2} {\left(  {b_1}^{2} +  {a_1}\right)} r^{2}\right]}  \Phi \nonumber \\
    &+ 8\sqrt{2}iE b_{1} r^{4}\,  \sqrt{{E^{2} r^{6} -L^{2}\left[  r^{4} - 2 \,  r^{3} + 8 \, {a_1} r - 16 \,  {a_1} - 2 \, r{\left( r^{2} - 4 \,  \right)} \epsilon\right]}}
\end{align}
Notice that setting $\epsilon=a_1=b_1=0$ reproduces the results for the Schwarzschild spacetime \cite{Frolov:2024olb},
\begin{align}
   \kappa_{Schw}= \frac{3\sqrt{2} i \, L^{2} \Phi}{2 \, r^{5}}.
\end{align}
Here, the function $\Phi=\Phi(r)$, introduced in Appendix \ref{app_null_tetrad}, is determined by the following differential equation
\begin{align}\label{phi_diff_equation}
    l^\alpha \partial_\alpha\Phi=\frac{d\Phi}{d\lambda} = \frac{E}{L B(r)}.
\end{align}

In the discussion below, we adopt the rescaling of parameters in terms of the energy $E$, following the prescription of \cite{Frolov:2024olb}, 
\begin{align}\label{rescaling_parametrized}
\lambda &\equiv E\lambda, \qquad L \equiv \dfrac{L}{E}, \qquad L_z \equiv \dfrac{L_z}{E}, \qquad \vb*{l}\equiv \dfrac{\vb*{l}}{E}, \qquad \vb*{w} \equiv \dfrac{\vb*{w}}{E^2}\\
\Phi &\equiv E\Phi, \qquad \mathcal{R}  \equiv \dfrac{\mathcal{R}}{E} = \pm\sqrt{\frac{r^2 - L^2N^2}{B^2}}, \qquad \Theta \equiv \frac{\Theta}{E} = \pm \sqrt{L^2\sin^2\theta - L_z^2},\end{align}
where $\Theta(\theta)$ and $\mathcal{R}(r)$ are defined in Eq. \eqref{eqRTheta}. Under this rescaling, combining the equations \eqref{w_kappa}, \eqref{kappa_parametrized}, \eqref{tetrad_m} and \eqref{tetrad_mbar}, the vector $\vb*{w}$ takes the form
\begin{align}\label{we2}
   \vb*{w} =\psi \vb*{e}_{2},
\end{align}
where $\vb*{e}_2$ is given in eq. \eqref{e1t_e2} and $\psi$ is given by
\begin{align}
    \psi = \frac{\mathcal{B}\sigma}{ \, L r^{4}{\left( r^{3} + 4 \,  {b_1} \right)^{3}}  \omega}.
\end{align}
In the above expression, $\sigma = \pm 1$ denotes the photon's helicity, $\omega$ is the frequency of the wave, and $\mathcal{B}$ is given by
\begin{align}\label{BB}
 \mathcal{B} = &    \left\{{ L^{3} \left[3   r^{5} - 20 r^{3} - 4 \,b_{1} r {\left(  r^{2} + 4 \right)} \right]}  \Phi \epsilon \right. \nonumber
    \\ & \left. + L{\left[16 L^{2} {b_1} r^{4} + 3  L^{2} r^{5} - 8   {b_1} r^{6} + 64  L^{2} {b_1}^{3} - 16  L^{2} {a_1} {b_1} r\right.}\right. \nonumber
    \\ & \left. {\left. + 64  L^{2} {a_1} {b_1} 
      - 4 L^{2}\, {\left(5   {a_1} +  {b_1}\right)} r^{3} + 48 L^{2}\, {\left(  {b_1}^{2} +  {a_1}\right)} r^{2}\right]}  \Phi \right. \nonumber
    \\ & \left. + 8 \, {b_1} r^{4} \sqrt{{ r^{6} -L^{2} \left[ r^{4} - 2 \,  r^{3} + 8 \,  {a_1} r - 16 \,  {a_1} - 2 \,r {\left( r^{2} - 4 \, \right)} \epsilon\right]}}\right\} .
\end{align}
{Again, assigning $\epsilon=a_1=b_1=0$ the results for Schwarzschild spacetime are recovered,
\begin{align}\label{psi_S}
    \psi_{Schw} = \frac{3L^{2}\sigma \Phi}{\omega r^{5}}.
\end{align}}

Since we are considering a spherically symmetric spacetime, we may assume, without loss of generality, that the unperturbed null vector $l^{\mu}$ lies in the equatorial plane. This allows us to set $ \theta = \pi/2$, yielding
\begin{align}\label{spherically_symmetric_condition}
   L=L_{z}, \qquad \Theta = 0, \qquad \vb*{l}= \frac{1}{N^2} \frac{\partial}{\partial t} + \frac{\mathcal{R}}{r} \frac{\partial}{\partial r}  + \frac{L}{r^{2}} \frac{\partial}{\partial \phi}, \qquad \vb*{e}_\theta \equiv \eval{\vb*{e}_{2}}_{\theta={\pi/2}} = \frac{1}{r}\frac{\partial}{\partial \theta}
\end{align}
so that the trajectory of the corresponding null ray is governed by the following differential equations
\begin{align}\label{EDO_geodesic_ray_parametrized}
    \frac{{\rm d}t}{{\rm d}\lambda} = \frac{1}{N^{2}}, \qquad \frac{{\rm d}r}{{\rm d}\lambda} = \frac{\mathcal{R}}{r}, \qquad \frac{{\rm d}\phi}{{\rm d}\lambda} = \frac{L}{r^{2}}.
\end{align}
Notice that, under the rescaling given in Eq. \eqref{rescaling_parametrized} and the assumption of a planar trajectory, $L$ coincides with the impact parameter. Assuming $L$ is sufficiently large to admit a region where the null ray reaches a minimum distance from the black hole before propagating back out to infinity, we have
\begin{align}
    \left.\frac{{\rm d}r}{{\rm d}\lambda}\right|_{\lambda=0}  = \frac{\mathcal{R}}{r_{m}}= 0.
\end{align}
This condition implies
\begin{align}
  L = \frac{r_{m}}{\sqrt{N^{2}(r_{m})}} = \frac{{r_m}^{3}}{\sqrt{{r_m}^{4} - {r_m}^{3} + {a_1} {r_m} - {\left({r_m}^{3} - {r_m}\right)} \epsilon - {a_1}}}.
\end{align}
Here, the affine parameter $\lambda$ is chosen such that $r(\lambda=0) = r_{m}$ is the radial coordinate at closest approach to the black hole, corresponding to the radius of the photon sphere.

To understand how the helicity–curvature interaction affects the trajectory of photons propagating in spacetime, we assume that their four-velocity vector can be written as
\begin{align}
    \ell^{\mu} = l^{\mu} + k^{\mu} = \frac{{\rm d}x^{\mu}}{{\rm d}\lambda} + \frac{{\rm d}\left(\delta x^{\mu}\right)}{{\rm d}\lambda},
\end{align}
where $l^{\mu}$ is the null geodesic vector describing the trajectory of the null ray in the absence of helicity-induced corrections. The vector $k^{\mu}$, on the other hand, accounts for the perturbation due to the interaction between the photon's helicity and the spacetime curvature, containing terms of order $\mathcal{O}\left(1/\omega\right)$. {According to the spinoptics formalism, $\vb*{\ell}$ must be a null vector. Thus, in our high-frequency approximation, $\ell^{\mu}\ell_{\mu} = \mathcal{O}(1/\omega^{2}) \approx 0$. It follows that $\vb*{k}$ and $\vb*{l}$ are mutually orthogonal. We have already constructed two vectors, $\vb*{\widetilde{e}}_1$ and $\vb*{e}_2$, orthogonal to $\vb*{l}$ (see Eq. \eqref{e1t_e2}). Therefore, $\vb*{k}$ can be expressed as a linear combination of $\vb*{\widetilde{e}}_1$ and $\vb*{e}_2$. In the coordinate basis on the equatorial plane, this becomes
\begin{align}
    \vb*{k} = k^1\left[\frac{{B(r)}\mathcal{R}(r)}{L N^2(r)}\,\frac{\partial}{\partial t }
    + \frac{r}{L {B(r)}}\,\frac{\partial}{\partial r}\right] \;+\; \frac{k^2}{r}\,\frac{\partial}{\partial\theta}.
\end{align}
In the following, it is convenient to write $\vb*{k}$ explicitly in the coordinate basis,
\begin{align}
    \vb*{k} = k^t\frac{\partial}{\partial t} + k^r\frac{\partial}{\partial r} + k^\theta\frac{\partial}{\partial\theta},
\end{align}
whose components are
\begin{align}
    k^t = k^1\frac{{B(r)}\mathcal{R}(r)}{L N^2(r)}, \qquad
    k^r = k^1\frac{r}{L {B(r)}}, \qquad
    k^\theta = \frac{k^2}{r}.
\end{align}
} From eqs.\eqref{eq_l_geodesic} and \eqref{we2} we can write
\begin{align}\label{desviation_diff_eq_general}
    D\ell^{\mu} = \frac{{\rm d}\ell^{\mu}}{{\rm d}\lambda} + \Gamma^{\mu}_{\phantom{\mu}\alpha\beta}\ell^{\alpha}\ell^{\beta} = \left(\frac{{\rm d}l^{\mu}}{{\rm d}\lambda} + \Gamma^{\mu}_{\phantom{\mu}\alpha\beta}l^{\alpha}l^{\beta}\right) + \left(\frac{{\rm d}k^{\mu}}{{\rm d}\lambda} + 2\Gamma^{\mu}_{\phantom{\mu}\alpha\beta}l^{\alpha}k^{\beta}\right) + \Gamma^{\mu}_{\phantom{\mu}\alpha\beta}k^{\alpha}k^{\beta} = \psi e_{\theta}^{\mu}.
\end{align}
The terms in the first parenthesis vanish because $\vb*{l}$ is a geodesic vector, whereas the terms after the second parenthesis are of order $\mathcal{O}(1/\omega^{2})$ and therefore also vanish in our approximation. Consequently, the only relevant equations are
\begin{align}
    \frac{{\rm d}k^{\mu}}{{\rm d}\lambda} + 2\Gamma^{\mu}_{\phantom{\mu}\alpha\beta}l^{\alpha}k^{\beta} = \psi e_{\theta}^{\mu}.
\end{align}
{This equation can be written in component form as
\begin{align}
         \frac{{\rm d} k^{t}}{{\rm d}\lambda} + k^{t}\frac{\mathcal{R}}{r}\frac{{\rm d} }{{\rm d}r}\ln N^{2}  + \frac{k^{r}}{N^{2}}\frac{{\rm d} }{{\rm d}r}\ln N^{2}  &= 0,   \\
         \frac{{\rm d} k^{r}}{{\rm d}\lambda} + k^{t}\frac{N^{2}}{B^{2}}\frac{\rm{d}}{\rm{d}r}N^{2}  + k^{r}\frac{\mathcal{R}}{r}\frac{\rm{d}}{\rm{d}r}\left[\ln\frac{B^{2}}{N^{2}}\right]  &= 0,\\
         \frac{{\rm d} k^{\theta}}{\rm d\lambda} + \frac{2\mathcal{R}}{r^{2}}k^{\theta} &= \frac{\psi}{r},\\ \label{kr0}
         \frac{2L k^{r}}{r^{3}} &= 0.
\end{align}
From Eq.~\eqref{kr0}, we have $k^r = 0$, which implies $k^1 = k^t = 0$. Consequently, the perturbation affects only the component orthogonal to the equatorial plane. The only remaining equation is
\begin{align}\label{spinoptic_equation_parametrized_black_hole}
    \frac{{\rm d}^{2}{\delta\tilde{\theta}}}{{\rm d}{\lambda}^{2}} + \frac{2{\mathcal{R}}}{{r}^{2}}\frac{{\rm d}\delta\tilde{\theta}}{{\rm d}{\lambda}} = \frac{\mathcal{B}}{ 3 \, L r^{5} \left(r^{2} + 4b_{1}\right)^{3}}.
\end{align}

Here we adopt the definition $\delta\tilde{\theta} = \delta\theta \, \omega/3\sigma$. The resulting equation governs the particle's deflection angle with respect to the equatorial plane. Notably, this angle depends on the particle's interaction with the spacetime curvature, which is ultimately determined by the black hole parameters $\epsilon$, $a_1$, and $b_{1}$. {Notice that, from eqs. \eqref{BB} and \eqref{psi_S}, it is straightforward to see that the results for the Schwarzschild geometry are reproduced when $\epsilon=a_1=b_1=0$.} 

We now perform a numerical integration of the system of differential equations defined by eqs. \eqref{phi_diff_equation}, \eqref{EDO_geodesic_ray_parametrized}, and \eqref{spinoptic_equation_parametrized_black_hole}, adopting the following initial conditions
\cite{Frolov:2024olb}:
\begin{align}\label{initial_conditions_general}
\left.\Phi(\lambda)\right|_{\lambda=0}=0,\qquad \left.r(\lambda)\right|_{\lambda=0}=r_m,\qquad \left.\phi(\lambda)\right|_{\lambda=0}=0,\qquad  \left.\dot{\theta}(\lambda)\right|_{\lambda=0}=\left.{\theta}(\lambda)\right|_{\lambda=0}=0.    
\end{align}
The numerical results from the integration are presented for various combinations of the Rezzolla–Zhidenko coefficients $a_1$ and $b_1$, as well as for different values of the relative deviation: $\epsilon = -0.1$ (dashed line), $\epsilon = 0.0$ (dash-dotted line), and $\epsilon = 0.1$ (dotted line). In Fig. \ref{th_parametrized_black_holes}, we plot the deflection angle $\delta\tilde{\theta}$ as a function of the affine parameter $\lambda$. The Schwarzschild solution (solid line) is included. In Fig. \ref{graf_A_th}, where $\epsilon = a_1 = b_1 = 0$, the dash-dotted line overlaps the solid one.

As expected, the deflection angle decreases with increasing distance from the black hole. The relative deviation $\epsilon$, conversely, causes the deflection angle to increase. This follows from the fact that $\epsilon$ measures the deviation of the Rezzolla–Zhidenko horizon from the Schwarzschild horizon (see Section \ref{sec3}). Positive $\epsilon$ values correspond to larger horizon radii, hence a stronger gravitational field. This enhanced gravity intensifies the photon's spin–curvature interaction, leading to a greater deflection from the equatorial plane. Notably, in Figs.~\ref{graf_B_th} and~\ref{graf_C_th}, the deflection angle $\delta\tilde{\theta}$ for $\epsilon = 0$ is closer to the Schwarzschild values, especially at $r_m = 3.5$ and $r_m = 4.0$. Nevertheless, the Rezzolla--Zhidenko black holes with $\epsilon = 0$, $a_1 = b_1 = \pm 0.1$ originate from beyond-GR theories, as $B^{2}(r) \neq 1$. A similar agreement in values is also observed for the quasinormal modes of these black holes, as reported in Ref. \cite{Dubinsky:2024rvf}.

\begin{figure}[H]
    \centering
    \begin{subfigure}{0.5\textwidth}
        \centering
        \includegraphics[width=\linewidth]{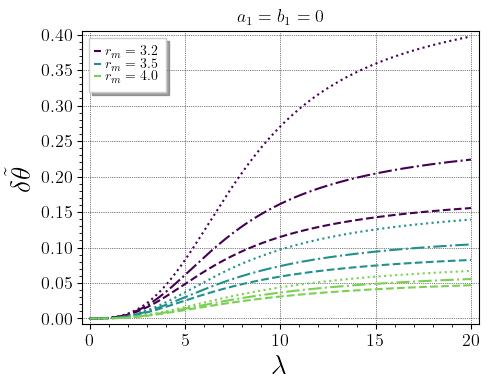}
        \caption{}
        \label{graf_A_th}
    \end{subfigure}%
    \begin{subfigure}{0.5\textwidth}
        \centering
        \includegraphics[width=\linewidth]{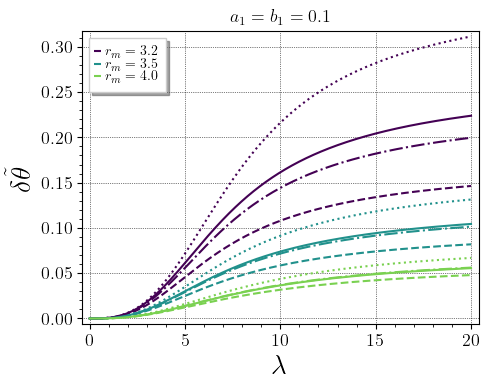}
        \caption{}
        \label{graf_B_th}
    \end{subfigure}

    \vspace{0.3cm}

    \begin{subfigure}{0.5\textwidth}
        \centering
        \includegraphics[width=\linewidth]{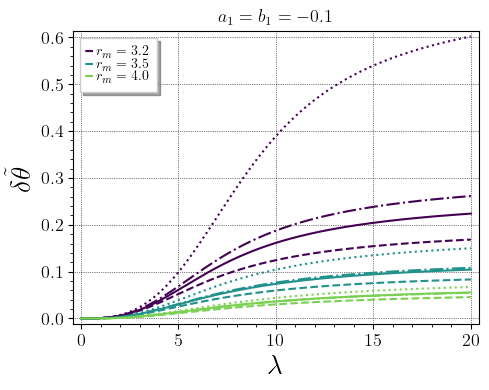}
        \caption{}
        \label{graf_C_th}
    \end{subfigure}%
    \begin{subfigure}{0.5\textwidth}
        \centering
        \includegraphics[width=\linewidth]{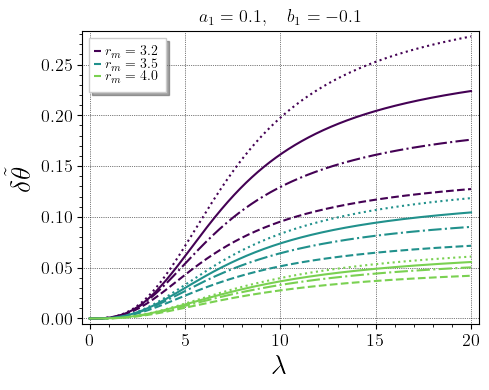}
        \caption{}
        \label{graf_D_th}
    \end{subfigure}

    \caption{\footnotesize {Normalized} deflection angle $\delta\tilde{\theta}= \delta\theta \, \omega / 3\sigma$ of a light ray with respect to the equatorial plane of the orbit for several values of the Rezzolla–Zhidenko parametrization coefficients $a_1$ and $b_1$. The relative deviation parameter is given by $\epsilon=-0.1$ (dashed line), $\epsilon=0.0$ (dash-dotted line), and $\epsilon=0.1$ (dotted line). The solid lines correspond to the deflection angle for the Schwarzschild solution. Note that in panel (a) the solid line was omitted, as it coincides with the dash-dotted line.}
    \label{th_parametrized_black_holes}
\end{figure}

In Fig.~\ref{ph_parametrized_black_holes}, we plot the polar angle $\phi$ as a function of the affine parameter $\lambda$ for the same Rezzolla--Zhidenko coefficients used for $\delta\tilde{\theta}$. A similar behavior observed for the equatorial deviations also appear for $\phi$. Moreover, we define $\delta\tilde{\theta} = \delta\theta \, \omega / 3\sigma$, so that photons with helicity $\sigma = 1$ experience a deflection $\delta\tilde{\theta}$, while those with $\sigma = -1$ experience $-\delta\tilde{\theta}$. This helicity-dependent separation of trajectories is known as gravitational birefringence~\cite{Murk:2024qgj}.

\begin{figure}[H]
    \centering
    \begin{subfigure}{0.5\textwidth}
        \centering
        \includegraphics[width=\linewidth]{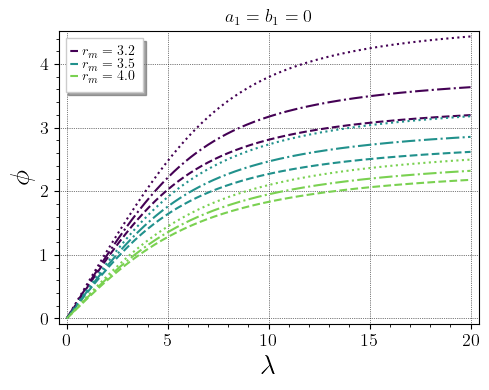}
        \caption{}
        \label{graf_ph_para}
    \end{subfigure}%
    \begin{subfigure}{0.5\textwidth}
        \centering
        \includegraphics[width=\linewidth]{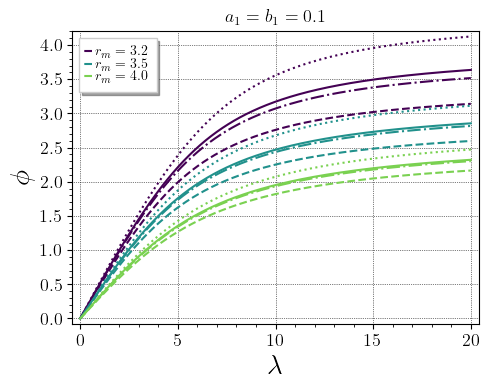}
        \caption{}
    \end{subfigure}

    \vspace{0.3cm}

    \begin{subfigure}{0.5\textwidth}
        \centering
        \includegraphics[width=\linewidth]{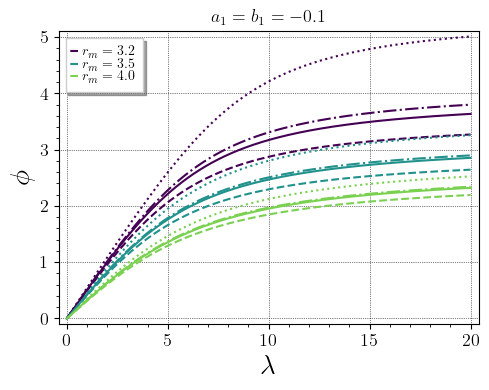}
        \caption{}
    \end{subfigure}%
    \begin{subfigure}{0.5\textwidth}
        \centering
        \includegraphics[width=\linewidth]{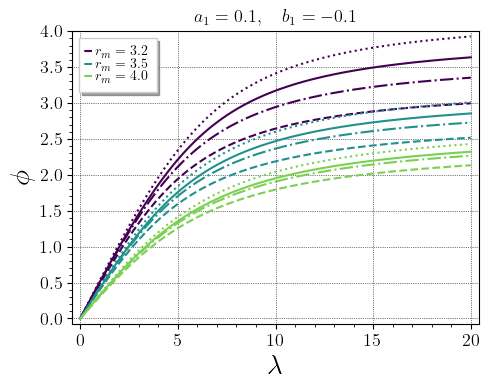}
        \caption{}
    \end{subfigure}

    \caption{\footnotesize $\phi-$angle as a function of $\lambda$ for several values of the Rezzolla–Zhidenko parametrization coefficients $a_1$ and $b_1$. The relative deviation parameter is given by $\epsilon=-0.1$ (dashed line), $\epsilon=0.0$ (dash-dotted line), and $\epsilon=0.1$ (dotted line). The solid lines correspond to the deflection angle for the Schwarzschild solution. Note that in panel (a) the solid line was omitted, as it coincides with the dash-dotted line.}
    \label{ph_parametrized_black_holes}
\end{figure}

\subsection{Hairy Black Holes}

We now apply the spinoptics approach to regular hairy black holes described by Eqs. \eqref{regular_metric_black_hole} and \eqref{functions_regular_metric}. For this spacetime, the null tetrad given in Eqs. \eqref{tetrad_l}–\eqref{tetrad_n} is obtained by setting $N^{2}(r) = f(r,\beta)$ and $B^{2}(r) = 1$, where $f(r,\beta)$ is the function defined in Eq. \eqref{functions_regular_metric}. Following the same procedure as for the previous parametrized metric, we can derive an expression for $\kappa$ in the case of a regular hairy black hole
\begin{align}
   \kappa = \frac{3  \, \sqrt{2} i L^{2}  \Phi}{2 \, r^{5}} -\frac{\, 3\sqrt{2} i L^{2} \, e^{\left(-\frac{r}{\beta }\right)}}{2} \, {\left(\frac{1 }{6\beta^{3} r^{2}} + \frac{1 }{2\beta^{2} r^{3}} + \frac{1  }{\beta r^{4}} + \frac{ 1 }{r^{5}}\right)}\Phi.  
\end{align}
Again, observe that in the limit
\begin{align}
   \lim_{\beta \to 0} \kappa = \frac{3  \, \sqrt{2} i L^{2}  \Phi}{2 \, r^{5}} = \kappa_{Schw},
\end{align}
the result obtained in \cite{Frolov:2024olb} for the Schwarzschild black hole is recovered. It is worth noting that the value of $\kappa$ for this solution is consistently lower than $\kappa_{Schw}$. This indicates that the presence of the additional hair parameter weakens the interaction between photon helicity and spacetime curvature, which follows directly from the attenuation of the curvature itself (see Figure \ref{fig:Kretschmann_Scalar}).

In the following discussion, we again rescale the parameters in terms of the energy and constrain the null radius vector $\vb*{l}$ to the equatorial plane. Under these conditions, the vector $w^\mu$ from Eq. \eqref{eq_l_geodesic} takes the form $w^{\mu} = \psi , e_{2}^{\mu}$, where $\psi$ is given by
\begin{align}
\psi = \frac{3 L^{2} \Phi \sigma}{r^{5} \omega} - \frac{3 L^{2} \sigma e^{-r/\beta}}{\omega} \left( \frac{1}{r^{5}} + \frac{1}{r^{4} \beta} + \frac{1}{2 r^{3} \beta^{2}} + \frac{1}{6 r^{2} \beta^{3}} \right) \Phi,
\end{align}
the $\vb*{l}$ components by
\begin{align}\label{EDO_geodesic_ray}
    \frac{{\rm d}t}{{\rm d}\lambda} = \frac{1}{f(r,\beta)}, \qquad \frac{{\rm d}r}{{\rm d}\lambda} = \frac{\mathcal{R}}{r}, \qquad \frac{{\rm d}\phi}{{\rm d}\lambda} = \frac{L}{r^2},
\end{align}
and the vector $e^{\mu}_{2}$ is adjusted according to the spacetime metric. Consequently, for the regular hairy black hole, the differential equation governing the particle's deflection angle relative to the equatorial plane becomes, 
\begin{align}\label{spinoptic_equation_regular_black_hole}
\frac{{\rm d}^{2}{\delta\tilde{\theta}}}{{\rm d}{\lambda}^{2}} + \frac{2{\mathcal{R}}}{{r}^{2}}\frac{{\rm d}\delta\tilde{\theta}}{{\rm d}{\lambda}} = \frac{L \lambda}{r^{6}} \left[ 1 - \left( 1 + \frac{r}{\beta} + \frac{r^{2}}{2 \beta^{2}} + \frac{r^{3}}{6 \beta^{3}} \right) e^{-r/\beta} \right].
\end{align}
{Adopting $\delta\tilde{\theta} = \delta\theta \, \omega / 3\sigma$ as before, and with Eq. \eqref{phi_diff_equation} reducing to $\dv{\Phi}{\lambda} = \frac{1}{L}$, $\Phi$ simplifies to $\Phi = \lambda / L$, in agreement with Ref. \cite{Frolov:2024olb}.} Note that this deflection angle depends on the particle's interaction with the spacetime curvature, which ultimately depends on the additional parameter $\beta$. 

\begin{figure}[h!]
    \centering
    \includegraphics[width=0.48\linewidth]{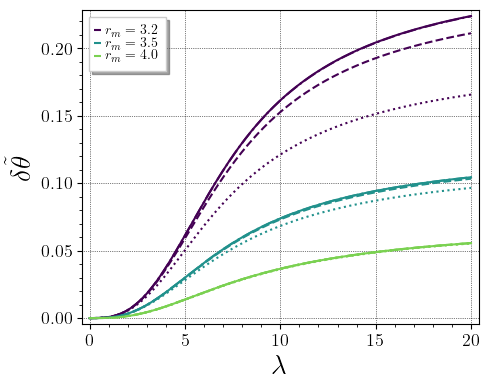}
    \includegraphics[width=0.48\linewidth]{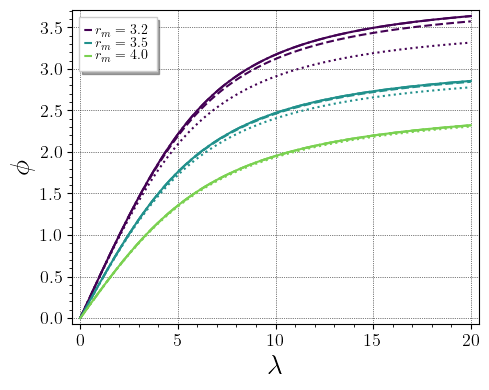}
    \caption{\footnotesize Left panel: {Normalized} deflection angle $\delta\tilde{\theta}= \delta\theta \, \omega / 3\sigma$ of a light ray with respect to the equatorial plane of the orbit for Schwarzschild (solid line), $\beta=0.2$ (dash-dotted line), $\beta=0.3$ (dashed line) and $\beta=\beta_{crit}$ (dotted). Note that the curves coincide for $r_m=4$. Right panel:  $\phi-$angle for Schwarzschild (solid line), $\beta=0.2$ (dashdot line), $\beta=0.3$ (dashed line) and $\beta=\beta_{crit}$ (dotted). Note that again the curves coincide for $r_m=4$.}
    \label{fig:th_ph_regular}
\end{figure}

We numerically integrated the system of differential equations given by Eqs.~\eqref{EDO_geodesic_ray} and \eqref{spinoptic_equation_regular_black_hole} using the same initial conditions as in Eq.~\eqref{initial_conditions_general}. The results are presented in Fig.~\ref{fig:th_ph_regular}. We observe in the left panel of Fig.~\ref{fig:th_ph_regular} that the deflection angle is attenuated both by the particle's distance from the black hole and by the presence of the additional parameter $\beta$. Notably, the influence of $\beta$ becomes more significant for particles closer to the black hole, where it acts to moderate the deflection angle. For $r_m = 3.2$, the case $\beta = 0.2$ coincides with the Schwarzschild result, rendering the effect of the additional parameter imperceptible, whereas deviations become apparent for $\beta = 0.3$ and $\beta = \beta_{\rm crit}$. At $r_m = 3.5$, both $\beta = 0.2$ and $\beta = 0.3$ converge to the Schwarzschild case. For $r_m = 4$, all deflection angles coincide with the Schwarzschild value regardless of $\beta$, so that the influence of the parameter becomes negligible. An analogous behavior is observed for the polar angle $\phi$, as shown in the right panel of Fig.~\ref{fig:th_ph_regular}.

Once we have calculated the deflection angle due to the helicity--curvature interaction for both the parametrized and the hairy regular black hole, we can further test the capability of the parametrized metric to reproduce the results of the hairy solution. This is shown in Fig.~\ref{relative_error_spinoptics} for several values of $\beta$. The results clarify what Table~\ref{tab:parameters_parametrized_black_hole} already suggests: the Rezzolla--Zhidenko (RZ) metric best approximates the hairy black hole as the hairy parameter $\beta$ approaches zero. For $\beta = 0.1$, the relative error is $E_R = \dfrac{|\delta\tilde{\theta}_{\rm hairy} - \delta\tilde{\theta}_{\rm RZ}|}{\delta\tilde{\theta}_{\rm hairy}} \sim 10^{-5}$, where $\delta\tilde{\theta}_{\rm hairy}$ and $\delta\tilde{\theta}_{\rm RZ}$ denote the off-equatorial-plane deflection angles for the hairy solution and the RZ metric, respectively. However, the relative error becomes extremely large as $\beta$ approaches $\beta_{\rm crit}$, reaching $E_R\sim 500\%$ for $\beta = 0.33$. This highlights the limitations of the RZ metric in reproducing the results of the hairy solution. {Increasing the number of parameters in the RZ metric could refine the above results. However, this would not only differ from what is suggested in the results of Ref.~\cite{Konoplya:2020hyk}, but would also render the parametrized metric impractical and computationally more expensive.}

\begin{figure}[H]
    \centering
    \begin{subfigure}{0.5\textwidth}
        \centering
        \includegraphics[width=\linewidth]{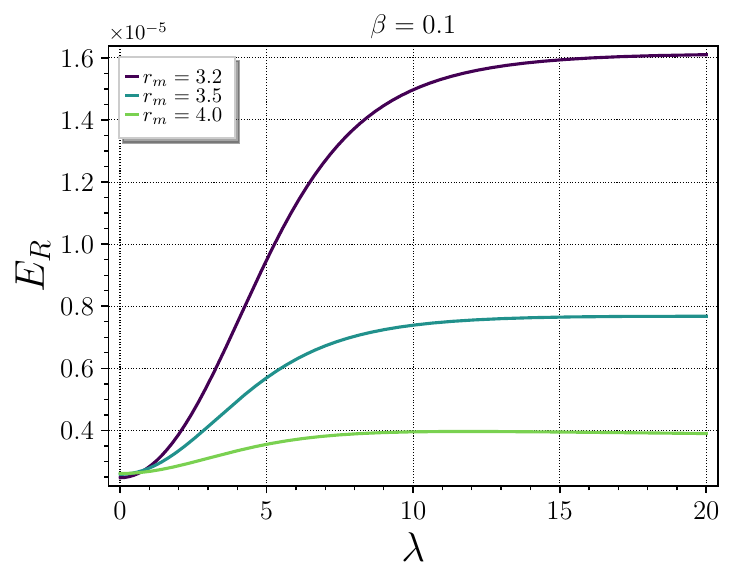}
        \caption{}
        \label{graf_ER_01}
    \end{subfigure}%
    \begin{subfigure}{0.5\textwidth}
        \centering
        \includegraphics[width=\linewidth]{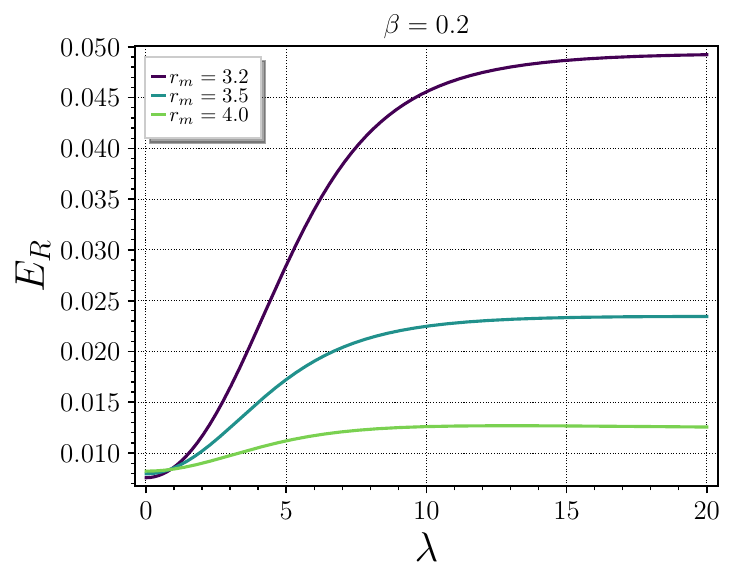}
        \caption{}
    \end{subfigure}

    \vspace{0.3cm}

    \begin{subfigure}{0.5\textwidth}
        \centering
        \includegraphics[width=\linewidth]{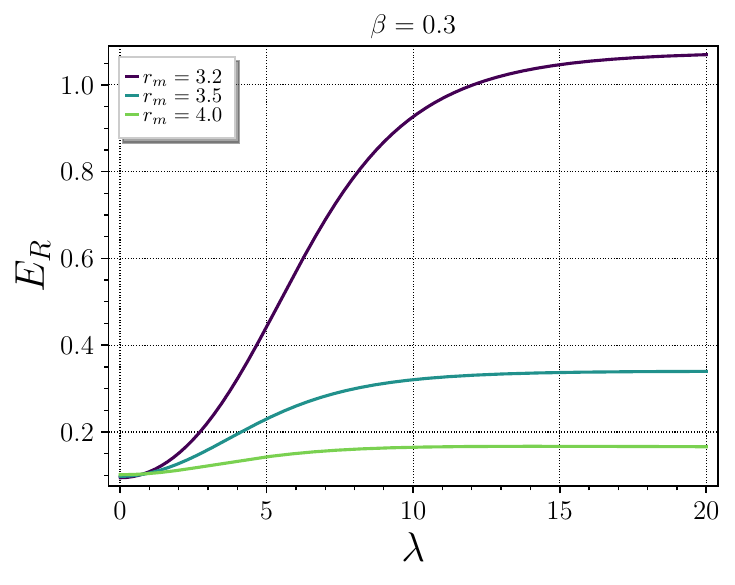}
        \caption{}
    \end{subfigure}%
    \begin{subfigure}{0.5\textwidth}
        \centering
        \includegraphics[width=\linewidth]{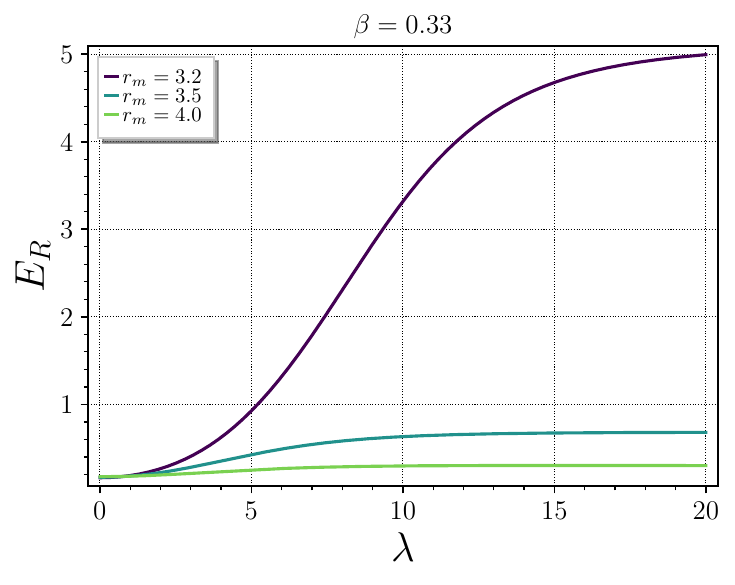}
        \caption{}
    \end{subfigure}

    \caption{\footnotesize Relative error $E_R = |\delta\tilde{\theta}_{\rm hairy} - \delta\tilde{\theta}_{\rm RZ}|/\delta\tilde{\theta}_{\rm hairy}$ between the deflection angles due to the helicity-curvature interaction. Here $\delta\tilde{\theta}_{\rm hairy}$ denotes the deflection angle for the regular hairy black hole and $\delta\tilde{\theta}_{\rm RZ}$ for its approximation via the RZ metric (see Section~\ref{sec3}), as a function of $\lambda$ for different values of $\beta$.}
    \label{relative_error_spinoptics}
\end{figure}

\begin{figure}[h!]
    \centering
    \includegraphics[width=0.48\linewidth]{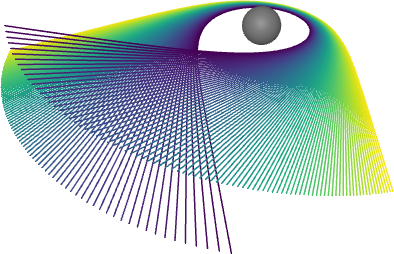}
    \includegraphics[width=0.45\linewidth]{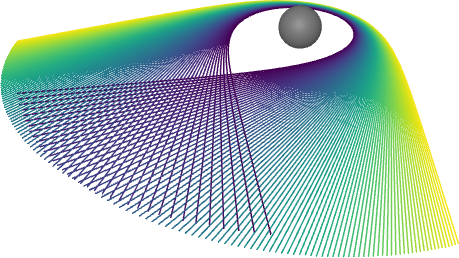}
    \includegraphics[width=0.48\linewidth]{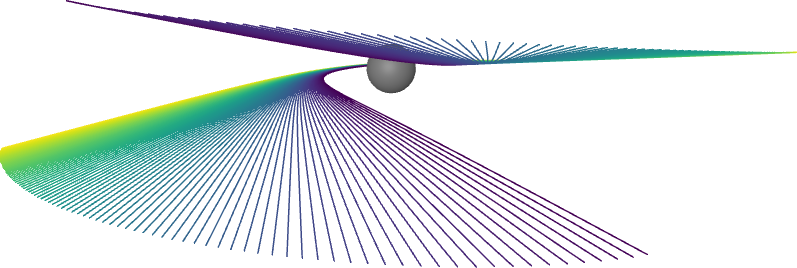}
    \includegraphics[width=0.45\linewidth]{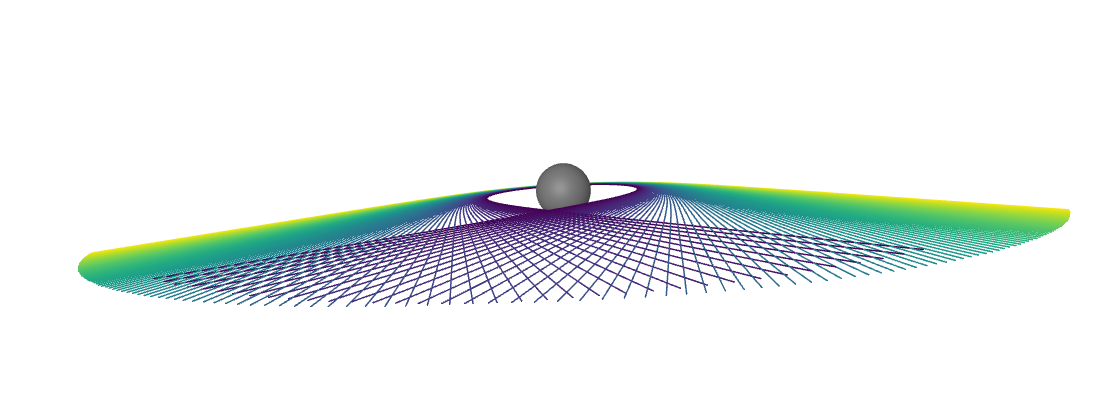}
    \caption{\footnotesize {Numerically calculated orbits of light rays in the vicinity of the regular hairy black hole, for $r_m$ ranging from 3.1 (dark purple) to 4 (yellow), with $\beta = 0.3883 \approx \beta_{\text{crit}}$, helicity $\sigma = 1$, and the grey sphere representing the event horizon for the corresponding $\beta$. The left panels show orbits including spinoptics corrections, while the right panels show those without corrections, which follow planar geodesics. Again, the normalized deflection angle $\delta\tilde{\theta}= \delta\theta \, \omega / 3\sigma$ was adopted to plot the trajectories.}
    }
    \label{orbits}
\end{figure}

In Fig.~\ref{orbits}, we show the numerically calculated\footnote{The symbolic and numerical calculations, as well as the plots, were performed using the open-source computer algebra system SageMath \cite{sagemath}, along with the SageManifolds \cite{Gourgoulhon:2014ywa} and SciPy \cite{2020SciPy-NMeth} projects.} orbits of light rays in the vicinity of the {regular hairy black hole, specified by Eqs. \eqref{regular_metric_black_hole} and \eqref{functions_regular_metric},} for $r_m$ ranging from 3.1 (dark purple) to 4 (yellow) with $\beta = 0.3883 \approx \beta_{\text{crit}}$. The orbits, including spinoptics corrections, are shown on the left, while those without corrections are shown on the right. This explicitly demonstrates that when the helicity--curvature interaction is taken into account, the light rays are no longer confined to the equatorial plane. As expected, the spinoptics effect becomes stronger for rays that approach the black hole more closely. {Notice, however, that as in previous plots we have used the normalized deflection angle $\delta\tilde{\theta}= \delta\theta \, \omega / 3\sigma$. Although this definition renders the effect of non-planar orbits explicit, it does not represent the actual deflection angle faithfully. The actual deflection angle scales inversely with frequency: $\delta\theta\sim \delta\tilde{\theta}/\omega$.} These modifications of the trajectories near the horizon may influence the black hole shadow image, as suggested in Ref.~\cite{Frolov:2025bva}.

\section{Concluding remarks}
\label{sec6}

In this paper, we investigated null ray dynamics in the presence of helicity-curvature interactions. To this end, we selected two models to describe the 
{background}. The first is the Rezzolla--Zhidenko parametrized metric \cite{Rezzolla:2014mua, DeLaurentis:2017dny, Konoplya:2020hyk}, which describes a wide class of spherically symmetric spacetimes, both within and beyond General Relativity. The second model is 
{a}
regular hairy black hole solution derived via the gravitational decoupling method \cite{Ovalle:2023ref, Ovalle:2017fgl}. {Before applying the spinoptics formalism to these spacetimes, we tested the ability of the Rezzolla--Zhidenko metric, at the order suggested in Ref.~\cite{Konoplya:2020hyk}, to describe the regular hairy black hole. Both metrics could be used to describe low-energy corrections to general relativity solutions. However, to the best of our knowledge, they had not been compared as we have done here.} %
%
In fact, the Rezzolla--Zhidenko model best approximates the hairy black hole as the hairy parameter $\beta$ approaches zero, that is, the closer the hairy black hole is to the Schwarzschild solution. This assessment is first carried out by comparing the shadow of the hairy black hole with that of the Rezzolla–Zhidenko metric when the latter is used to describe the same hairy black hole, following \cite{Konoplya:2020hyk}. The relative error reaches $\sim 6.7940\%$ as $\beta$ approaches $\beta_{\rm crit}\approx 0.3883945$.

We proceed by applying the spinoptics formalism to the spacetime models mentioned above. The main outcome of this formalism for our purpose is that, due to the helicity--curvature interaction, the propagation of light rays
in spherically symmetric spacetimes is not planar. To achieve this, the results of Ref.~\cite{Frolov:2024olb} 
were extended 
to generalized spherically symmetric spacetimes where $g_{rr}g_{tt}\neq -1$. The off-equatorial-plane deflection angles 
were then computed
for various configurations of the RZ metric, as well as for a range of the hairy parameter $\beta$ for the hairy regular solution, and compared with the results for the Schwarzschild solution. Our results show a clear signature of the additional parameters in the deflection angle, which might lead to a reexamination of black hole shadow images, as suggested in Ref.~\cite{Frolov:2025bva}. Moreover, the RZ metric 
was used here to reproduce the results of the hairy regular solution and to compare them with those obtained directly from the latter. {This highlights the limitations of the RZ metric, at the order employed, in mimicking the hairy black hole.} In particular, the relative error between them becomes unacceptable as $\beta$ approaches $\beta_{\rm crit}$, reaching $\sim 500\%$ for $\beta = 0.33$. {While extending the RZ metric could yield some improvement, Ref.~\cite{Konoplya:2020hyk} suggests that this improvement is limited, and would come at the cost of increased computational expense.}

\appendix

\section{Parallel-propagated null tetrad}
\label{app_null_tetrad}

In this appendix, we derive the null tetrad that is parallel-propagated along null geodesics for a general spherically symmetric metric of the form given by the RZ metric:
\begin{align}\label{RZ_metric_app}
{\rm d}s^{2} = - N^{2}(r) {\rm d}t^{2} + \frac{B^{2}(r)}{N^{2}(r)}{\rm d}r^{2} + r^{2} {\rm d}\theta^{2} + r^{2}\sin^{2}\theta {\rm d}\phi^{2}.
\end{align}
We follow the approach of Refs. \cite{Frolov:2024ebe, Frolov:2017kze}, which in summary proceeds as follows: starting from the null tangent vector, it uses the orthogonal projection properties of the spherically symmetric conformal Killing–Yano and Killing–Yano forms to obtain two additional mutually orthogonal vectors. It then finds a fourth vector orthogonal to the plane formed by these two. From this set of four vectors, the tetrad basis parallel-propagated along null geodesics is constructed.

Let $\displaystyle \vb*{\xi} = \pdv{t}$ and $\displaystyle \vb*{\zeta} = \pdv{\phi}$ denote the Killing vectors associated with local energy conservation and angular momentum in the $z$-direction, respectively. The closed conformal Killing–Yano form $\vb{h}$ is given as the solution to the equation 
\begin{align}\label{CKY}
\nabla_\lambda h_{\mu\nu} &= \xi_\mu g_{\nu\lambda} - \xi_\nu g_{\mu\lambda}, 
\end{align}
and is dual to the Killing–Yano form $\vb{k} = \star \,\vb{h}$. The difficulty here is that $\vb{h} = -r {\rm d} t \wedge {\rm d} r$ is not a solution to the above equation, as in Ref. \cite{Frolov:2024ebe}. In fact, assuming a general antisymmetric two-form $\vb{h}(r,\theta)$ and substituting it into the equation, we find that only the component $h_{01}(r)$ is not identically zero. The only relevant equations in \eqref{CKY} are those corresponding to the components $t\theta\theta$ and $trr$, whose solutions are respectively given by  $h_{01} = r g_{tt} g_{rr} = -r B(r)^2$ and $g_{tt} g_{rr} = c$,
where $c$ is an integration constant. Together, they imply that $B(r)$ should be constant and $h_{01}$ proportional to $r$. Since this case is not of interest here, we disregard the $trr$ component and retain a more general expression for $h_{01}$, namely $\widetilde{h}_{01} = -r F(g_{tt}, g_{rr})$, where $F$ is a general function of the metric components $g_{tt}$ and $g_{rr}$.  To determine $F$, we first note that the Killing–Yano form from Ref. \cite{Frolov:2024ebe},
\begin{align}
\vb{k} = r^3 \sin\theta \, {\rm d}\theta \wedge {\rm d}\phi,
\end{align}
still satisfies $\nabla_{(\lambda} k_{\mu)\nu} = 0$, and therefore remains a Killing–Yano form. Consequently, $K_{\mu\nu} = k_{\mu}^{\ \alpha} k_{\nu\alpha}$
is a symmetric rank-2 Killing tensor.  We then fix $F$ by imposing that $\widetilde{\vb{h}} = -\star \vb{k} = -r B(r) \, {\rm d}t \wedge {\rm d}r$,
which implies 
\begin{align}
\widetilde{h}_{01} = -r F(g_{tt}, g_{rr}) = -r \sqrt{-g_{tt} g_{rr}}.
\end{align}
Since $\widetilde{h}^{\mu\alpha} k_{\alpha\nu}$ vanishes, it follows that for any vector ${v}$, the three vectors $k^{\mu}{_\nu} v^\nu$, $\widetilde{h}^{\mu}{_\nu} v^\nu$, and ${v}$ are mutually orthogonal. The construction of the parallel-propagated null tetrad below follows the same general procedure as in Ref. \cite{Frolov:2024ebe}. However, certain key details must be modified. Accordingly, we present it here for the sake of clarity and completeness.

Given a null geodesic $x^\mu(\lambda)$, its tangent vector can be expressed in terms of conserved quantities. This tangent vector also allows us to construct two independent vectors orthogonal to it by applying $\vb{k}$ and $\vb{\widetilde{h}}$ as above. Considering metric \eqref{RZ_metric_app}, the tangent vector is
\begin{align}
\vb*{l} =\frac{{\rm d}x^\mu}{{\rm d}\lambda}\frac{\partial}{\partial x^\mu }= \frac{E}{N^2(r)} \frac{\partial}{\partial t } + \frac{\mathcal{R}(r)}{r} \frac{\partial}{\partial r } + \frac{\Theta(\theta)}{r^{2} \sin\theta} \frac{\partial}{\partial \theta } + \frac{L_{z}}{r^{2} \sin^{2}\theta} \frac{\partial}{\partial \phi },
\end{align}
where $-\xi_\alpha l^\alpha = E,\quad \zeta_\alpha l^\alpha = L_z,\quad K_{\mu\nu} l^\mu l^\nu = L^2$ correspond, respectively, to the energy, the azimuthal angular momentum, and the total angular momentum. The functions $\mathcal{R}(r)$ and $\Theta(\theta)$ are given by
\begin{align}\label{eqRTheta}
 \mathcal{R}(r) = \pm\sqrt{\frac{E^{2} r^{2} - L^{2} N^2(r)}{B^2(r)}},\quad 
\Theta(\theta) = \pm\sqrt{L^{2} \sin^{2}\theta - L_{z}^{2}}.
\end{align}
The signs $\pm$ in eq. \eqref{eqRTheta} are independent. They are positive when the corresponding coordinate increases along the light ray's propagation and negative otherwise. Sign changes occur at the turning points, where either $\mathcal{R}$ or $\Theta$ vanishes. From these, the two normalized vectors orthogonal to ${l}$ are given by ${\widetilde{e}}_1^\mu = \dfrac{1}{L}\,\widetilde{h}^\mu{_\nu}\,l^\nu$ and ${e}_2^\mu = \dfrac{1}{L}\,k^\mu{_\nu}\,l^\nu.$ Explicitly, they read 
\begin{align}\label{e1t_e2}
\vb*{\widetilde{e}}_1 =  \frac{{B(r)}\mathcal{R}(r)}{L N^2(r)}\,\frac{\partial}{\partial t }
+ \frac{E r}{L {B(r)}}\,\frac{\partial}{\partial r}\; \quad\text{and}\quad \;\displaystyle \vb*{e}_2 = \frac{L_{z}}{L r \sin\theta}\,\frac{\partial}{\partial\theta }
- \frac{\Theta(\theta)}{L r \sin^{2}\theta}\,\frac{\partial}{\partial\phi}.    
\end{align}

The remaining independent vector, $\vb*{\widetilde{e}}_3$, can be obtained by starting with an arbitrary vector $\vb*{u}$ and imposing orthogonality to the plane spanned by $\vb*{\widetilde{e}}_1$ and $\vb*{e}_2$. This condition does not determine $\vb*{u}$ uniquely; it fixes only two of its components. Uniqueness is achieved by additionally requiring that $\vb*{u}$ be normalized and future-directed, satisfying $u^\mu l_\mu = -1$. The resulting vector is then denoted by $\vb*{\widetilde{e}}_3$, and is explicitly given by 
\begin{align}
\displaystyle \vb*{\widetilde{e}}_3 = \frac{E r^{2}}{2 \, L^{2} N^2(r)} \frac{\partial}{\partial t } + \frac{r R(r)}{2 \, L^{2}} \frac{\partial}{\partial r } -\frac{\Theta(\theta)}{2 \, L^{2} \sin\theta} \frac{\partial}{\partial \theta } -\frac{L_{z}}{2 \, L^{2} \sin^{2} \theta} \frac{\partial}{\partial \phi}.
\end{align}

The final step is to construct, from $\vb*{l}$, $\vb*{\widetilde{e}}_1$, $\vb*{e}_2$, and $\vb*{\widetilde{e}}_3$, a complex null tetrad that is parallel-propagated along the null geodesic. By definition, $\vb*{l}$ is parallel-propagated along the null geodesic. Explicit computation shows that $\vb*{e}_2$ is also parallel-propagated, i.e., $\nabla_{\vb*{l}}\vb*{e}_2 = 0$. For $\vb*{\widetilde{e}}_1$, notice that $\nabla_{\vb*{l}}\vb*{\widetilde{e}}_1 = -\dfrac{\xi_\alpha l^\alpha}{LB(r)} \vb*{l}= \dfrac{E}{LB(r)} \vb*{l}$, which suggests that we may subtract an appropriate multiple of $\vb*{l}$ from $\vb*{\widetilde{e}}_1$ to obtain a vector that is parallel-propagated along the null geodesic. To this end, we introduce a scalar field $\Phi(r)$ such that, along the geodesic, 
\begin{align}
\nabla_l(\Phi(r) \vb*{l}) = l^\alpha\partial_\alpha \Phi(r) \vb*{l} = -\dfrac{\xi_\alpha l^\alpha}{L{B}(r)} \vb*{l}.
\end{align}
Consequently, $\nabla_{\vb*{l}}(\vb*{\widetilde{e}}_1 - \Phi(r) \vb*{l}) = \nabla_{\vb*{l}}\vb*{\widetilde{e}}_1 - \nabla_{\vb*{l}}(\Phi(r) \vb*{l}) = 0$. Hence, by construction, the vector $\vb*{e}_1 = \vb*{\widetilde{e}}_1 - \Phi(r) \vb*{l}$ is parallel-propagated along the null geodesic. Observing that $\nabla_{\vb*{l}}\vb*{\widetilde{e}}_3 = \dfrac{E}{LB(r)}\vb*{\widetilde{e}}_1$, we apply the same strategy. Defining $n = \vb*{\widetilde{e}}_3 - \Phi(r)\vb*{\widetilde{e}}_1 + \dfrac{\Phi(r)^2}{2}\vb*{l}$ one finds that $\vb*{n}$ is likewise parallel-propagated by construction. The result is a set of parallel-propagated vectors from which we construct the following complex null tetrad
\begin{align}
\begin{cases}
\vb*{l}&= \vb*{l},\\
\vb*{m} &=\frac{1}{\sqrt{2}} (\vb*{e}_1 + i\vb*{e}_2) =  \frac{1}{\sqrt{2}} (\vb*{\widetilde{e}}_1 - \Phi(r) \vb*{l}+ i\vb*{e}_2),\\
\vb*{\bar{m}} &= \frac{1}{\sqrt{2}} (\vb*{e}_1 - i\vb*{e}_2) = \frac{1}{\sqrt{2}} (\vb*{\widetilde{e}}_1 - \Phi(r) \vb*{l}- i\vb*{e}_2),\\
\vb*{n} &= \vb*{\widetilde{e}}_3 - \Phi\vb*{\widetilde{e}}_1 + \dfrac{\Phi^2}{2}\vb*{l}, 
\end{cases}
\end{align}
with ${l}^\mu n_{\mu} = m^\mu m_\mu = -1$. Or, expressed in the coordinate basis
\begin{align}\label{tetrad_l}
\vb*{l}&= \frac{E}{N^2(r)} \frac{\partial}{\partial t } + \frac{\mathcal{R}(r)}{r} \frac{\partial}{\partial r } + \frac{\Theta(\theta)}{r^{2} \sin\theta} \frac{\partial}{\partial \theta } + \frac{L_{z}}{r^{2} \sin^{2}\theta} \frac{\partial}{\partial \phi }\\[10pt] 
\vb*{m} &= -\frac{1}{\sqrt{2}}\left[\frac{ E L \Phi(r) - B(r) R(r)}{ L N^2(r)} \frac{\partial}{\partial t }  +\frac{L B(r) R(r) \Phi(r) - E r^{2}}{ L r B(r)} \frac{\partial}{\partial r } +\frac{L \Theta(\theta) \Phi(r) - i \, L_{z} r}{ L r^{2} \sin\theta} \frac{\partial}{\partial \theta }\right. \notag \\ \label{tetrad_m}
&\qquad\qquad   \left.+\frac{L L_{z} \Phi(r) + i \, r \Theta(\theta)}{ L r^{2} \sin^2\theta} \frac{\partial}{\partial \phi }\right]\\[10pt]
\vb*{\bar{m}} &= -\frac{1}{\sqrt{2}}\left[\frac{ E L \Phi(r) - B(r) R(r)}{ L N^2(r)} \frac{\partial}{\partial t }  +\frac{L B(r) R(r) \Phi(r) - E r^{2}}{ L r B(r)} \frac{\partial}{\partial r } +\frac{L \Theta(\theta) \Phi(r) + i \, L_{z} r}{ L r^{2} \sin\theta} \frac{\partial}{\partial \theta }\right. \notag \\ \label{tetrad_mbar}
&\qquad\qquad   \left.+\frac{L L_{z} \Phi(r) - i \, r \Theta(\theta)}{ L r^{2} \sin^2\theta} \frac{\partial}{\partial \phi }\right]\\[10pt]
\vb*{n} &= \frac{1}{2 \, L^{2}}\left[\frac{ (L^{2} \Phi(r)^{2}+r^2)E - 2 \, L B(r) R(r) \Phi(r)}{ N^2(r)}  \frac{\partial}{\partial t } + \frac{(L^{2} \Phi(r)^{2} + r^{2} ) R(r)B(r) - 2 \, E L r^{2} \Phi(r)}{r B(r)}  \frac{\partial}{\partial r } \right.\notag \\
&\qquad\qquad + \left.\frac{{(L^{2} \Phi(r)^{2} - r^{2})} \Theta(\theta)}{r^{2} \sin\theta} \frac{\partial}{\partial \theta } +  \frac{(L^{2}  \Phi(r)^{2} - r^{2}) L_{z}}{r^{2} \sin^{2}\theta} \frac{\partial}{\partial \phi }\right]. \label{tetrad_n}
\end{align}
This completes the construction of a null tetrad that is parallel-propagated along the null ray in a general spherically symmetric spacetime.

\subsection*{\textbf{Acknowledgements}}
 KSA thanks to Coordenação de Aperfeiçoamento de Pessoal de Nível Superior – Brasil (CAPES) –- Finance Code 001, and the Rio de Janeiro State University (UERJ) for their hospitality.
 RTC thanks the National Council for Scientific and Technological Development - CNPq (Grant No. 401567/2023-0) for partial financial support. 

\section*{References}
\bibliography{references}

\end{document}